\IfFileExists{IEEEtran.cls}{%
  \documentclass[journal,12pt,onecolumn,draftclsnofoot]{IEEEtran}%
}{%
  \documentclass[12pt,letterpaper,onecolumn]{article}%
  \usepackage[margin=1in]{geometry}%
  \usepackage{mathptmx}%
}
\usepackage[numbers,sort&compress]{natbib}
\usepackage{amsmath,amssymb,amsfonts,amsthm,mathtools,bm}
\usepackage{graphicx}
\usepackage{float}
\usepackage{booktabs}
\usepackage{array}
\usepackage{enumitem}
\usepackage{needspace}
\usepackage{microtype}
\usepackage{xcolor}
\usepackage[colorlinks=true,linkcolor=blue,citecolor=blue,urlcolor=blue]{hyperref}
\usepackage{pgfplots}
\pgfplotsset{compat=1.18}

\ifdefined\pdfmapfile
  \pdfmapfile{+ps2pk35.map}
  \pdfmapfile{+cm.map}
  \pdfmapfile{+cmextra.map}
  \pdfmapfile{+symbols.map}
  \pdfmapfile{+rsfs.map}
\fi

\providecommand{\IEEEauthorblockN}[1]{#1}
\providecommand{\IEEEauthorblockA}[1]{\\#1}
\providecommand{\IEEEPARstart}[2]{\noindent\textbf{\MakeUppercase{#1}}#2}
\makeatletter
\@ifundefined{IEEEkeywords}{%
  \newenvironment{IEEEkeywords}{\begin{quote}\small\textbf{Index Terms---}}{\end{quote}}%
}{}
\makeatother

\hypersetup{
  pdftitle={Information-Theoretic Generalization for Adaptive Learning Under Privacy and Fairness Constraints},
  pdfauthor={Amirreza Zamani, Parastoo Sadeghi, Mikael Skoglund}
}

\setlist{nosep,leftmargin=*}

\newtheorem{theorem}{Theorem}
\newtheorem{lemma}{Lemma}
\newtheorem{proposition}{Proposition}
\newtheorem{corollary}{Corollary}
\theoremstyle{definition}
\newtheorem{definition}{Definition}
\newtheorem{example}{Example}
\theoremstyle{remark}
\newtheorem{remark}{Remark}

\newcommand{\E}{\mathbb{E}}
\newcommand{\Prob}{\mathbb{P}}
\newcommand{\R}{\mathbb{R}}
\newcommand{\cA}{\mathcal{A}}
\newcommand{\cS}{\mathcal{S}}
\newcommand{\cX}{\mathcal{X}}
\newcommand{\cY}{\mathcal{Y}}
\newcommand{\cT}{\mathcal{T}}
\newcommand{\cW}{\mathcal{W}}
\newcommand{\cU}{\mathcal{U}}
\newcommand{\gen}{\operatorname{gen}}

\newcommand{\ind}{\mathbf{1}}
\newcommand{\hfun}{\mathsf{h}}
\newcommand{\hfair}{\mathsf{h}^{\mathrm{fair}}}
\newcommand{\pos}[1]{\left[#1\right]^+}

\title{Information-Theoretic Generalization for Adaptive Learning Under
Privacy and Fairness Constraints}
\author{
\IEEEauthorblockN{Amirreza Zamani, Parastoo Sadeghi $^\dagger$, Mikael Skoglund 
		\IEEEauthorblockA{\\
		    \normalsize Department of Information Science and Engineering, KTH Royal Institute of Technology \\
			$^\dagger$School of Engineering and Technology, UNSW, Canberra, Australia\\
			Email: \protect amizam@kth.se, p.sadeghi@unsw.edu.au, skoglund@kth.se }}
            \vspace{-5mm}
            }
\newcolumntype{L}[1]{>{\raggedright\arraybackslash}p{#1}}
\newtheorem{appendixlemma}{Lemma}[section]
\providecommand{\appendices}{\appendix}
\begin{document}
\maketitle
\begin{abstract}
Adaptive learners may update, select, and stop models after repeatedly
examining the same data.  Privacy and fairness constraints regulate this
adaptation only selectively: they limit information about a private or
sensitive coordinate, whereas task generalization depends on information
about the coordinate on which loss is evaluated. We develop a unified
information-theoretic framework for this mismatch.  For a finite-alphabet pair
$(A,B)$, we define directly the residual-information certificate
$
\hfun_\epsilon(P_{AB})=\min\{H(B),H(B| A)+\epsilon\}.
$
A chain-rule transfer inequality converts a bound on
$I(W;A)$ into an upper bound on $I(W;B)$, which can be inserted into a
mutual-information generalization inequality.  We first establish a
transcript-level composition theorem for sequential learners.  Conditional
per-update information budgets add along the transcript, and the resulting
certificate remains valid for history-dependent checkpoint selection,
bounded stopping, and independent heterogeneous data batches.
Under selective privacy, the learner observes
$(X^n,Y^n)$, protects $X^n$, and incurs loss on $Y^n$.  Under fairness, the
learner observes $(S^n,X^n,T^n)$, limits dependence on the sensitive block
$S^n$, compresses $X^n$, and incurs loss on $T^n$.  The bounds expose
the residual entropies $H(Y| X)$, $H(T| S)$, and $H(T| X)$ and the
cumulative information cost of adaptation.  We retain block and
individual-sample bounds, task-compression refinements, per-output
total-variation privacy, empirical-to-population
demographic-parity transfer, and finite-alphabet information-parity transfer.
FRL/EFRL/SFRL/ESFRL constructions study the tightness of the task-information
upper bounds.
\end{abstract}

\begin{IEEEkeywords}
Adaptive learning, algorithmic fairness, demographic parity, generalization
bounds, information theory, mutual information, privacy, representation
learning.
\end{IEEEkeywords}

\section{Introduction}
\label{sec:introduction}

\IEEEPARstart{I}{nformation-theoretic} generalization theory explains why a
data-dependent hypothesis can perform similarly on a fresh population and on
the sample that produced it.  The question becomes more delicate when a model
is updated repeatedly.  Hyperparameters may be chosen from earlier
checkpoints, a stopping rule may depend on observed performance, and a final
model may be selected after many interactions with the same data.  A
canonical statement says that, for a sub-Gaussian loss, the expected
generalization error is controlled by the mutual information between the
released hypothesis and the task sample
\cite{russo2016controlling,xu2017information}:
\begin{equation}
 \left|\E[\gen(W,Z^n)]\right|
 \leq \sqrt{\frac{2\sigma^2}{n}I(W;Z^n)}.             \label{eq:intro-classic}
\end{equation}
This principle has led to global, conditional, and individual-sample
information bounds and to connections with adaptive data analysis,
algorithmic stability, iterative learning, and differential privacy
\cite{dwork2015generalization,bassily2016algorithmic,
steinke2020reasoning,bu2020tightening}.
The information constraint available in privacy- or fairness-constrained
learning is often different from the information term in
\eqref{eq:intro-classic}.  Suppose one observation is $Z=(X,Y)$, where $X$ is
private and $Y$ carries task information.  Selective privacy constrains
\begin{equation}
 I(W;X^n)\leq\epsilon_n,                               \label{eq:intro-privacy}
\end{equation}
whereas a loss depending on $Y$ is controlled by $I(W;Y^n)$.  Likewise, for
$Z=(S,X,T)$, bounded demographic parity can constrain
\begin{equation}
 I(W;S^n)\leq\epsilon_n,                               \label{eq:intro-fair}
\end{equation}
while task generalization depends on $I(W;T^n)$.  Neither
\eqref{eq:intro-privacy} nor \eqref{eq:intro-fair} can simply be substituted for
$I(W;Z^n)$ in \eqref{eq:intro-classic}.  The protected coordinate and the loss
coordinate are different random objects.

This mismatch is not merely notational.  If $Y$ has randomness that remains
after $X$ is known, a learner can retain information about that residual
component without paying for it through $I(W;X^n)$.  In the fairness model,
bounded dependence on $S^n$ controls task information only to the extent that
$S$ explains $T$.  Feature compression supplies a second route when $X$
explains $T$.  Thus a selective constraint leads naturally to two terms: a
budget divided by sample size and a conditional-entropy residual that need not
decay with $n$.
Sequential adaptation adds another mismatch.  A guarantee stated only for
one checkpoint need not account for information exposed by earlier updates,
model selection, or a stopping decision.  We therefore treat the complete
interaction record as a transcript.  If update $k$ reveals at most
$\epsilon_{n,k}$ conditional nats about a protected block, the transcript
reveals at most $\sum_k\epsilon_{n,k}$ nats by the chain rule.  A terminal
model, a history-selected checkpoint, and a bounded stopping rule are all
post-processings of that transcript.  This gives an auditable information
ledger for continually updated models without requiring independent updates
or a schedule fixed in advance.

Privacy--utility and fairness--utility studies formalize related mechanism
design tradeoffs in terms of retained information under leakage and rate
constraints \cite{makhdoumi2014privacy,zamani2024privacy,
zamani2026fairness}.  For a particular learner $W$, the chain rule gives
\begin{equation}
 I(W;B)\leq \min\{H(B),H(B\mid A)+I(W;A)\}.           \label{eq:intro-transfer}
\end{equation}
This observation motivates a closed-form function
\begin{equation}
 \hfun_\epsilon(P_{AB})
 :=\min\{H(B),H(B\mid A)+\epsilon\},                  \label{eq:intro-h}
\end{equation}
which supplies the upper-bound certificate used throughout the paper.  The
same function covers both selective privacy and bounded demographic parity.
In the sequential setting, the one-shot budget is replaced by the cumulative
conditional budget of the transcript.\\
\textbf{Positioning and novelty:}
The paper sits at the intersection of four literatures that are usually
developed separately.  First, mutual-information generalization starts with a
dependence measure involving the task sample or the complete data
\cite{russo2016controlling,xu2017information,bu2020tightening}.  Second,
adaptive data analysis controls the cumulative effect of repeated,
history-dependent access to data
\cite{dwork2015generalization,bassily2016algorithmic,steinke2020reasoning}.
Third, privacy mechanism design studies how much useful information can be disclosed
while limiting leakage about a secret \cite{makhdoumi2014privacy,
zamani2024privacy}.  Fourth, fair representation learning controls dependence
between a representation and a sensitive attribute while preserving task
information \cite{zemel2013learning,hardt2016equality,
calmon2017optimized,zamani2026fairness}.
The novelty is the architecture connecting these lines.  The first layer is
an explicit information-transfer bridge from a \emph{selective} constraint
to the task information appearing in a generalization theorem.  The second
layer lifts that bridge from a one-shot learner to an adaptive transcript
through conditional information accounting.  The third layer separates risk
generalization from generalization of a measured fairness constraint.  The
result is a common proof system for terminal models, selected checkpoints,
and continually updated models under privacy or fairness budgets.  It also
distinguishes universal upper certificates from mechanism-design
achievability.  A nonvanishing certificate is inconclusive for a particular
algorithm; a construction showing large admissible task information is an
existence statement, not a lower bound on every learner's generalization
error.
The fairness part makes a second distinction that is frequently blurred.
Constraining information in the released learner output and transferring an
empirical demographic-parity statistic to the population are different
problems.  The first is controlled by $I(W;S^n)$ and the relation between $S$
and $T$; the second needs an adaptive-complexity condition involving the
training coordinates on which the fairness statistic is evaluated.  We give
separate theorems for both.
Relative to a static treatment, the transcript formulation is substantive:
privacy and fairness budgets compose conditionally across updates; arbitrary
history-dependent checkpoint selection is absorbed by data processing; and a
heterogeneous-stage theorem permits independent batches with different source
laws and sub-Gaussian parameters.  Relative to the two component models, the
paper uses one notation and proof architecture, employs a single closed-form
residual-information certificate, retains block, compression, coordinatewise,
and adaptive refinements, and separates upper bounds, tightness diagnostics, and
population-fairness transfer.\\
\textbf{Contributions:}
The main contributions are summarized as follows.
 I. We define the residual-information upper-bound function
 $\hfun_\epsilon$ in \eqref{eq:intro-h}, establish its structural properties,
 and prove a block information-transfer lemma valid for every admissible
 learner.
II. We prove a sequential composition theorem for arbitrary adaptive
 transcripts and extend it to independent, non-identically distributed
 batches.  Conditional information budgets add across updates, while the
 heterogeneous certificate combines stagewise residual entropies with a
 variance-weighted information term.  The results cover checkpoint selection
 and bounded early stopping.
III. For selective privacy, we prove expected generalization bounds under
 $I(W;X^n)\leq\epsilon_n$.  We derive task and full-sample compression
 refinements, individual-sample and total-variation variants, and their
 adaptive-transcript forms.
IV. For bounded demographic parity, we combine
 $I(W;S^n)\leq\epsilon_n$ with optional feature compression
 $I(W;X^n)\leq r_n$.  The active term is the minimum of the entropy cap and
 the two residual-plus-budget branches.  We establish adaptive and
 heterogeneous versions with checkpoint selection.
V. We separately analyze generalization of fairness statistics.  Under
 $I(W;S^n,X^n)\leq\kappa_n$, empirical demographic-parity gaps transfer to
 the population, and finite-alphabet mutual-information parity transfers via
 total variation and entropy continuity.  Conditional complexity budgets
 extend both statements to adaptive checkpoint selection.
VI. We use EFRL/ESFRL constructions to study tightness of the
 task-information upper bounds and support the theory with deterministic,
 noisy, independent-residual, binary-cascade, ternary, and balanced-group
 examples.\\  
\textbf{Organization:}
Section~\ref{sec:related} reviews related work.  Section~\ref{sec:prelim} develops the
one-shot and sequential machinery, including heterogeneous batches.  Part~I,
Sections~\ref{sec:privacy-model}, \ref{sec:privacy-main}, \ref{sec:privacy-refinements}, \ref{sec:privacy-adaptive}, \ref{sec:privacy-witness}, and~\ref{sec:privacy-examples},
treats selective privacy.  Part~II,
Sections~\ref{sec:fair-model}, \ref{sec:fair-main}, \ref{sec:fair-adaptive}, \ref{sec:fair-witness}, \ref{sec:fair-metrics}, and~\ref{sec:fair-examples},
treats bounded demographic parity and fairness-metric transfer.
Section~\ref{sec:discussion} collects design implications, robustness
interpretations, and limitations.  Detailed proofs are placed in the
appendices.

\section{Related Work}
\label{sec:related}
\textbf{Information-theoretic generalization:}
Russo and Zou \cite{russo2016controlling} bounded selection bias in adaptive
analysis using mutual information.  Xu and Raginsky
\cite{xu2017information} developed an information-theoretic analysis of
learning algorithms based on transportation and change-of-measure arguments.
Individual-sample bounds refine the global dependence measure by averaging
coordinatewise information quantities \cite{bu2020tightening}.  Conditional
mutual information supplies a distribution-independent framework based on a
supersample and selector variables \cite{steinke2020reasoning}.  Related
stability and reusable-holdout results include
\cite{dwork2015generalization,bassily2016algorithmic}.
These results also motivate transcript-level accounting: the output of an
adaptive procedure is the complete random object that mediates later
decisions, not merely the final numerical parameter.  Our setting differs
because the stated constraint controls neither the whole sample nor, in
general, the task coordinate.  We use a standard expected-generalization
lemma as a module, transfer protected information to task information, and
then compose the transferred constraint across history-dependent updates.
The heterogeneous-stage result additionally allows independent batches with
different source laws; it is not an adversarial-shift theorem.\\
\textbf{Selective privacy and privacy--utility tradeoffs:}
The privacy funnel and related formulations quantify disclosure utility under
information leakage constraints \cite{makhdoumi2014privacy}.  The observable-
and hidden-private-data models in \cite{zamani2024privacy} provide detailed
bounds and constructions under mutual-information and per-output privacy
criteria.  Functional representation
lemmas and their strong variants supply constructive descriptions of random
variables with prescribed independence or leakage properties
\cite{li2018strong,11686056}.
We borrow the semantic split between private and useful variables, while the
upper-bound function in \eqref{eq:intro-h} supplies the transfer certificate
used here.
For completeness, later sections translate construction results into
existence witnesses.  Those witnesses study tightness but are not needed
to establish any generalization upper bound.\\
\textbf{Fair representation learning:}
Fair representation learning seeks useful encodings whose distributions are
insensitive to protected attributes \cite{zemel2013learning}.  Demographic
parity, equality of opportunity, adversarial representation learning, and
preprocessing methods provide different operational approaches
\cite{hardt2016equality,calmon2017optimized,edwards2016censoring}.  Information-
theoretic formulations connect fairness to bottleneck and privacy concepts.
The bounded-demographic-parity model in \cite{zamani2026fairness} adds an
explicit encoding-rate constraint and derives constructive tradeoff bounds.
The present fairness analysis has two levels.  At the representation level,
we study how bounded information about $S^n$ controls task-risk
generalization.  At the statistical level, we study whether an empirical
fairness metric computed after adaptive training transfers to a fresh
population.  The second question requires complexity control not supplied by
demographic parity alone.

\section{Common Framework}
\label{sec:prelim}

\subsection{Notation and sampling assumptions}

Random variables are denoted by capital letters, their realizations by lower-
case letters, and alphabets by calligraphic letters.  For
$B^n=(B_1,\ldots,B_n)$, the notation $P_B^{\otimes n}$ denotes an i.i.d.
product distribution.  All logarithms are natural, so entropy and mutual
information are measured in nats.  The positive part is
$[a]^+=\max\{a,0\}$.  Unless stated otherwise, the protected and task
variables take values in finite alphabets and all displayed entropies are
finite.

For a loss $\ell:\cW\times\mathcal B\to\R$, define
\begin{align}
 L_P(w):=\E_{B\sim P_B}[\ell(w,B)], \
 L_n(w):=\frac1n\sum_{i=1}^n\ell(w,B_i),\
 \gen(W,B^n):=L_P(W)-L_n(W).                         \label{eq:gen-def}
\end{align}

\begin{definition}[Sub-Gaussian loss]
\label{def:subg}
The loss is $\sigma^2$-sub-Gaussian under $P_B$ if, for every fixed
$w\in\cW$ and every $\lambda\in\R$,
\begin{equation}
 \log\E\exp\!\left(\lambda[\ell(w,B)-L_P(w)]\right)
 \leq\frac{\lambda^2\sigma^2}{2}.                    \label{eq:subg}
\end{equation}
If $\ell\in[a,b]$, Hoeffding's lemma permits
$\sigma^2=(b-a)^2/4$.
\end{definition}

\begin{lemma}[Task-information generalization]
\label{lem:mi-generalization}
Let $B^n\sim P_B^{\otimes n}$, and let $W$ be any randomized output jointly
distributed with $B^n$.  Under \eqref{eq:subg},
\begin{equation}
 \left|\E[\gen(W,B^n)]\right|
 \leq\sqrt{\frac{2\sigma^2}{n}I(W;B^n)}.             \label{eq:mi-gen}
\end{equation}
Moreover,
 $\E\left|\gen(W,B^n)\right|
 \leq\sqrt{\frac{2\sigma^2}{n}
 \left(I(W;B^n)+\log2\right)}.$               
\end{lemma}

The first inequality is used for expected task risk.  The second, stronger
form controls an expected absolute deviation and is used for empirical-
fairness transfer.  Proofs are given in Appendix~\ref{app:mi-lemmas}.

\subsection[Residual-information upper-bound function]{The residual-information upper-bound function $\hfun_\epsilon$}

Let $(A,B)\sim P_{AB}$, where $A$ denotes a protected coordinate and $B$ a
task coordinate.

\begin{definition}[Residual-information upper-bound function]
\label{def:h}
For $\epsilon\geq0$, define
\begin{equation}
 \hfun_\epsilon(P_{AB})
 :=\min\{H(B),H(B\mid A)+\epsilon\}.                \label{eq:h-def}
\end{equation}
For an arbitrary block law $P_{A^nB^n}$, define
 $\hfun_{\epsilon_n}^{(n)}(P_{A^nB^n})
 :=\min\{H(B^n),H(B^n\mid A^n)+\epsilon_n\}.$  
\end{definition}
For an i.i.d. pair,
\begin{equation}
 \hfun_{\epsilon_n}^{(n)}(P_{AB}^{\otimes n})
 =n\hfun_{\epsilon_n/n}(P_{AB}).                      \label{eq:h-tensor}
\end{equation}

\begin{proposition}[Properties of $\hfun_\epsilon$]
\label{prop:h-properties}
For fixed $P_{AB}$, the map $\epsilon\mapsto\hfun_\epsilon(P_{AB})$ is
nondecreasing, concave, continuous, and one-Lipschitz.  It satisfies
\begin{align}
 \hfun_0(P_{AB})&=H(B\mid A),                         \label{eq:h-zero}\\
 \hfun_\epsilon(P_{AB})&=H(B),\quad
 \epsilon\geq I(A;B).                                \label{eq:h-sat}
\end{align}
The value in \eqref{eq:h-zero} is an upper-bound value and is not asserted to
be achievable by a zero-leakage release.
\end{proposition}

\begin{proof}
The function is affine with slope one until it reaches the constant $H(B)$,
after which its slope is zero.  The junction occurs at
$I(A;B)$.
\end{proof}

\begin{lemma}[Selective information transfer]
\label{lem:transfer}
For every jointly distributed $(U,A,B)$ with finite-entropy $B$,
\begin{equation}
 I(U;A)\leq\epsilon
 \quad\Longrightarrow\quad
 I(U;B)\leq\hfun_\epsilon(P_{AB}).                    \label{eq:transfer}
\end{equation}
The corresponding block statement holds with
$(A,B,\epsilon)$ replaced by $(A^n,B^n,\epsilon_n)$.
\end{lemma}

\begin{proof}
The proof follows by \cite{zamani2024privacy}. The chain rule gives
$ I(U;B)
 =I(U;A)+I(U;B\mid A)-I(U;A\mid B) 
 \leq I(U;A)+H(B\mid A) 
 \leq\epsilon+H(B\mid A).$                    
Combining this with $I(U;B)\leq H(B)$ gives
\eqref{eq:transfer}.
\end{proof}

\begin{remark}[Certificate and tightness]
\label{rem:h-scope}
The function $\hfun_\epsilon$ is a uniform certificate: every output meeting
the constraint is below it.  A nonvanishing value does not imply that a given
learner has large generalization error.  Construction-based witness results
later in the paper study whether admissible outputs can attain or approach
the task-information certificate.  Thus, their tightness statements concern
the information upper bound and do not by themselves give lower bounds on
the loss generalization error.
\end{remark}

\subsection{Adaptive transcripts and conditional information ledgers}
\label{sec:adaptive-common}

We now lift the direct certificate from one release to a sequential learner.
Fix a deterministic horizon $K$.  At update $k$, the learner produces a
random object $V_k$ according to an arbitrary kernel
\begin{equation}
 P_{V_k\mid D_n,\Pi_{k-1}},\qquad
 \Pi_k:=(V_1,\ldots,V_k),\quad \Pi_0:=\varnothing,  \label{eq:transcript-model}
\end{equation}
where $D_n=(A^n,B^n)$.  The object $V_k$ may contain a model checkpoint, a
query, a hyperparameter choice, an evaluation, or a stopping indicator.
Thus the entire history visible to later updates is part of $\Pi_k$.  A
released model $\widehat W_k$ is any randomized post-processing of $\Pi_k$;
the additional random seed is required to be independent of the data given
the transcript.

For nonnegative update budgets $\epsilon_{n,k}$ and $\rho_{n,k}$, consider
\begin{align}
 I(V_k;A^n\mid\Pi_{k-1})&\leq\epsilon_{n,k},          \label{eq:seq-a-budget}\\
 I(V_k;B^n\mid\Pi_{k-1})&\leq\rho_{n,k}.             \label{eq:seq-b-budget}
\end{align}
Write
 $\epsilon_{n,[k]}:=\sum_{j=1}^k\epsilon_{n,j},
 $ and $
 \rho_{n,[k]}:=\sum_{j=1}^k\rho_{n,j}.             $
The first ledger controls selective leakage; the optional second ledger
controls direct task memorization.  Update kernels may be selected from the
past transcript, while the displayed numbers are deterministic upper bounds
on the resulting conditional mutual informations.  No independence among
updates is assumed.

\begin{theorem}[Adaptive selective-transfer principle]
\label{thm:adaptive-transfer}
Let $(A_i,B_i)_{i=1}^n$ be i.i.d., let the loss on $B$ be
$\sigma^2$-sub-Gaussian, and let $\widehat W_k$ be any post-processing of
$\Pi_k$.  Under~\eqref{eq:seq-a-budget}, for every $k\leq K$,
\begin{align}
 I(\widehat W_k;B^n)
 &\leq\min\{nH(B),nH(B\mid A)+\epsilon_{n,[k]}\},   \label{eq:seq-transfer-info}\\
 |\E[\gen(\widehat W_k,B^n)]|
 &\leq\sqrt{2\sigma^2\min\left\{
 H(B),H(B\mid A)+\frac{\epsilon_{n,[k]}}n
 \right\}}.                                         \label{eq:seq-transfer-gen}
\end{align}
If~\eqref{eq:seq-b-budget} also holds, add
$\rho_{n,[k]}$ to the minimum in~\eqref{eq:seq-transfer-info} and
$\rho_{n,[k]}/n$ to the minimum in~\eqref{eq:seq-transfer-gen}.

Moreover, let $J\in\{1,\ldots,K\}$ be selected from the full transcript,
possibly with independent randomization, and suppose the released object
$(J,\widehat W_J)$ is a post-processing of $\Pi_K$.  Then the same bounds
hold with $k=K$ and $\widehat W_k$ replaced by $(J,\widehat W_J)$.
\end{theorem}

\begin{proof}
The mutual-information chain rule and~\eqref{eq:seq-a-budget} give
 $I(\Pi_k;A^n)=\sum_{j=1}^k I(V_j;A^n\mid\Pi_{j-1})
 \leq\epsilon_{n,[k]}.                        $
Applying Lemma~\ref{lem:transfer} to $(\Pi_k,A^n,B^n)$ and using data processing
for $\widehat W_k$ proves~\eqref{eq:seq-transfer-info}.  Similarly,
$I(\Pi_k;B^n)\leq\rho_{n,[k]}$ under~\eqref{eq:seq-b-budget}.  Intersect the
available bounds and apply Lemma~\ref{lem:mi-generalization}.  The selected
checkpoint statement follows because $(J,\widehat W_J)$ is itself a
post-processing of $\Pi_K$.
\end{proof}

\begin{remark}[Early stopping and what must be charged]
\label{rem:stopping}
A stopping time $\tau\leq K$ is covered by padding the transcript with a
fixed null symbol after stopping and releasing $(\tau,\widehat W_\tau)$.
If the decision to continue examines the data beyond what is already in the
transcript, that decision is included in $V_k$ and its conditional
information is charged.  A ``free'' stopping rule is justified only when it
is a post-processing of previously charged releases and public randomness.
\end{remark}

\subsection{Independent heterogeneous data arrivals}
\label{sec:heterogeneous-common}

Continual learning may also receive data in stages whose laws differ.  Let
\begin{equation}
 D^{(j)}=(A_j^{n_j},B_j^{n_j})\sim
 P_{A_jB_j}^{\otimes n_j},\qquad j=1,\ldots,m,       \label{eq:heterogeneous-data}
\end{equation}
with independence across stages, and put $N:=\sum_{j=1}^m n_j$.  The
learner may update on prefixes or on the full aggregate; all such behavior is
represented by~\eqref{eq:transcript-model}.  Define
\begin{align}
 L_{\rm mix}(w)&:=\frac1N\sum_{j=1}^m n_j
     \E_{B\sim P_{B_j}}[\ell(w,B)],\
 L_N(w):=\frac1N\sum_{j=1}^m\sum_{i=1}^{n_j}
     \ell(w,B_{j,i}),\nonumber\\
 \gen_{\rm het}(w)&:=L_{\rm mix}(w)-L_N(w).         \label{eq:heterogeneous-gen}
\end{align}

\begin{theorem}[Heterogeneous-stage adaptive bound]
\label{thm:heterogeneous-adaptive}
Suppose $\ell(w,B_j)$ is $\sigma_j^2$-sub-Gaussian under $P_{B_j}$ for every
fixed $w$.  Assume the conditional ledger~\eqref{eq:seq-a-budget} is imposed
with $A^n$ replaced by the aggregate protected data
$A^{[m]}=(A_1^{n_1},\ldots,A_m^{n_m})$.  Let
\begin{align}
 \mathsf H_B:=\sum_{j=1}^m n_jH(B_j),\
 \mathsf H_{B\mid A}:=\sum_{j=1}^m n_jH(B_j\mid A_j),\
 \mathsf V:=\sum_{j=1}^m n_j\sigma_j^2.            \label{eq:heterogeneous-sums}
\end{align}
For every transcript-selected terminal model $\widehat W$,
\begin{equation*}
 |\E[\gen_{\rm het}(\widehat W)]|
 \leq\frac1N\sqrt{2\mathsf V
 \min\{\mathsf H_B,\mathsf H_{B\mid A}+\epsilon_{N,[K]}\}}.
\end{equation*}
If the task-information ledger~\eqref{eq:seq-b-budget} holds for the
aggregate task data, the additional branch $\rho_{N,[K]}$ enters the minimum.
\end{theorem}
The proof is given in Appendix~\ref{app:adaptive-proofs}.
The theorem allows
different stage laws and loss scales, but it does not claim adversarial
distribution-shift robustness: the stages must be independent and the risk
is the sample-size-weighted mixture in~\eqref{eq:heterogeneous-gen}.

\begin{table*}[t]
\centering
\caption{Models, assumptions, and information-transfer substitutions.}
\label{tab:models}
\footnotesize
\begin{tabular}{@{}L{0.11\textwidth}L{0.17\textwidth}L{0.19\textwidth}L{0.19\textwidth}L{0.22\textwidth}@{}}
\toprule
Model & Observation and learner & Primary constraint & Task information & Direct certificate \\
\midrule
Selective privacy & $(X,Y)$; $P_{W\mid X^n,Y^n}$ & $I(W;X^n)\leq\epsilon_n$ & $I(W;Y^n)$ & $n\hfun_{\epsilon_n/n}(P_{XY})$ \\
Fairness & $(S,X,T)$; $P_{W\mid S^n,X^n,T^n}$ & $I(W;S^n)\leq\epsilon_n$ & $I(W;T^n)$ & $n\hfun_{\epsilon_n/n}(P_{ST})$ \\
Fairness with compression & same & additionally $I(W;X^n)\leq r_n$ & $I(W;T^n)$ &
$\begin{aligned}n\min\{&\hfun_{\epsilon_n/n}(P_{ST}),\\
&\hfun_{r_n/n}(P_{XT})\}\end{aligned}$ \\
Adaptive selective learning & transcript $\Pi_K$ &
$I(V_k;A^n\mid\Pi_{k-1})\leq\epsilon_{n,k}$ & $I(\Pi_K;B^n)$ &
$n\hfun_{\epsilon_{n,[K]}/n}(P_{AB})$ \\
Adaptive fairness & transcript $\Pi_K$ & conditional ledgers for $S^n$ and $X^n$ & $I(\Pi_K;T^n)$ & minimum of the two cumulative residual-plus-budget branches \\
Fairness-metric transfer & $(S,X)$ and $A_W(X)$ & $I(W;S^n,X^n)\leq\kappa_n$ & adaptive empirical process & absolute form of Lemma~\ref{lem:mi-generalization} \\
\bottomrule
\end{tabular}
\end{table*}

\section{Part I: Selective-Privacy Model}
\label{sec:privacy-model}

\subsection{Data, learner, and risk}

Let $(X,Y)\sim P_{XY}$ be a pair of discrete random variables on finite
alphabets $\cX$ and $\cY$.  The variable $X$ is private and $Y$ is
task-relevant.  The training sample is
\begin{equation}
 D_n=(X^n,Y^n)=\{(X_i,Y_i)\}_{i=1}^n
 \sim P_{XY}^{\otimes n}.                              \label{eq:privacy-data}
\end{equation}
A randomized learner is any channel $P_{W\mid X^n,Y^n}$ with
$W\in\cW$.  The loss $\ell:\cW\times\cY\to\R$ is evaluated on the task
coordinate and is $\sigma^2$-sub-Gaussian under $P_Y$ for every fixed $w$.
Thus
\begin{align}
 L_P(w)=\E_{Y\sim P_Y}[\ell(w,Y)],\
 L_n(w)=\frac1n\sum_{i=1}^n\ell(w,Y_i).              \label{eq:privacy-risk}
\end{align}

\begin{definition}[Selective mutual-information privacy]
\label{def:selective-privacy}
The learner is $(X^n,\epsilon_n)$-selectively private if
\begin{equation}
 I(W;X^n)\leq\epsilon_n.                               \label{eq:selective-privacy}
\end{equation}
\end{definition}

The constraint is distribution dependent.  It controls average information
leakage about the private block and is not, without additional assumptions, a
differential-privacy statement.  It also does not directly control
$I(W;Y^n)$ or $I(W;X^n,Y^n)$.

\subsection{Why a full-sample bound cannot be inserted}

The learner can depend on both coordinates.  In particular,
\begin{equation}
 I(W;X^n)\leq\epsilon_n
 \ \not\!\!\Longrightarrow\ 
 I(W;X^n,Y^n)\leq\epsilon_n.
\end{equation}
Even data processing does not compare
$I(W;X^n)$ and $I(W;Y^n)$ without a Markov structure.  The applicable
identity is instead
 $I(W;Y^n)
 =I(W;X^n)+I(W;Y^n\mid X^n)-I(W;X^n\mid Y^n),$
which exposes the uncontrolled conditional term
$I(W;Y^n\mid X^n)$.  Bounding that term by $H(Y^n\mid X^n)$ produces the
residual entropy in the results below.

\section{Selective-Privacy Bounds}
\label{sec:privacy-main}

\subsection{Block information and expected risk}

\begin{theorem}[Selective-privacy generalization]
\label{thm:privacy-main}
Under \eqref{eq:privacy-data}, \eqref{eq:selective-privacy} and the sub-Gaussian loss
assumption,
\begin{align}
 \left|\E[\gen(W,Y^n)]\right|
 &\leq\sqrt{\frac{2\sigma^2}{n}
 \hfun_{\epsilon_n}^{(n)}(P_{XY}^{\otimes n})}        \label{eq:privacy-main-block}\\
 &=\sqrt{2\sigma^2\min\!\left\{
 H(Y),H(Y\mid X)+\frac{\epsilon_n}{n}
 \right\}}.                                          \label{eq:privacy-main-explicit}
\end{align}
\end{theorem}

\begin{proof}
By Lemma \ref{lem:transfer},
\begin{equation}
 I(W;Y^n)\leq
 \hfun_{\epsilon_n}^{(n)}(P_{XY}^{\otimes n}).        \label{eq:privacy-task-info}
\end{equation}
Insert \eqref{eq:privacy-task-info} into \eqref{eq:mi-gen} and use the tensor
identity \eqref{eq:h-tensor}.
\end{proof}

The entropy cap $H(Y)$ matters when the privacy budget is large.  In the
nontrivial regime $\epsilon_n/n<I(X;Y)$, the active term is
$H(Y\mid X)+\epsilon_n/n$.  For a loss in $[a,b]$,
 $\left|\E[\gen]\right|
 \leq\frac{b-a}{\sqrt2}
 \sqrt{\min\!\left\{H(Y),H(Y\mid X)+\frac{\epsilon_n}{n}\right\}}.$
\begin{remark}[Residual uncertainty]
\label{rem:privacy-residual}
The term $H(Y\mid X)$ measures task randomness not accounted for by the
protected coordinate.  If it is positive and fixed, the theorem does not by
itself certify a vanishing error when $\epsilon_n=o(n)$.  This is a limitation
of the privacy-only certificate, not a claim that every learner overfits.
\end{remark}

\begin{corollary}[Deterministic task coordinate]
\label{cor:privacy-deterministic}
If $H(Y\mid X)=0$, then
\begin{align}
 I(W;Y^n)&\leq\min\{nH(Y),\epsilon_n\},\\
 \left|\E[\gen]\right|
 &\leq\sqrt{2\sigma^2
 \min\!\left\{H(Y),\frac{\epsilon_n}{n}\right\}}.    \label{eq:privacy-deterministic}
\end{align}
In particular, $\epsilon_n=o(n)$ is sufficient for the upper bound to vanish.
\end{corollary}

\subsection{Task and full-sample compression}

Selective privacy may be supplemented with a direct budget on task information:
\begin{equation}
 I(W;Y^n)\leq\rho_n.                                  \label{eq:task-compression}
\end{equation}
The stronger full-sample condition
$I(W;X^n,Y^n)\leq\rho_n$ implies \eqref{eq:task-compression} by data
processing.

\begin{theorem}[Privacy with compression]
\label{thm:privacy-compression}
If \eqref{eq:selective-privacy} and~\eqref{eq:task-compression} hold, then
\begin{equation}
 \left|\E[\gen]\right|
 \leq\sqrt{\frac{2\sigma^2}{n}
 \min\!\left\{\rho_n,nH(Y),nH(Y\mid X)+\epsilon_n\right\}}.
 \label{eq:privacy-compression-bound}
\end{equation}
The same bound holds under the stronger full-sample condition
\begin{equation}
 I(W;X^n,Y^n)\leq\rho_n.
\end{equation}
\end{theorem}

\begin{proof}
Combine \eqref{eq:privacy-task-info} with
$I(W;Y^n)\leq\rho_n$ and apply Lemma~\ref{lem:mi-generalization}.
\end{proof}

If $\rho_n=o(n)$, the bound in
\eqref{eq:privacy-compression-bound} vanishes irrespective of $H(Y\mid X)$.
For $\rho_n=O(\log n)$, its order is
$\sqrt{\log n/n}$.  Privacy controls sensitive memorization, whereas
compression controls residual task memorization.

\section{Refined Selective-Privacy Results}
\label{sec:privacy-refinements}

\subsection{Individual-sample information}

The global information $I(W;Y^n)$ can be pessimistic when the learner depends
unevenly on sample coordinates.  The individual-sample bound
\cite{bu2020tightening} gives
\begin{equation}
 \left|\E[\gen(W,Y^n)]\right|
 \leq\frac1n\sum_{i=1}^n\sqrt{2\sigma^2I(W;Y_i)}.
 \label{eq:privacy-ismi}
\end{equation}

\begin{theorem}[Individual selective privacy]
\label{thm:individual-privacy}
If $I(W;X_i)\leq\epsilon_i$ for $i=1,\ldots,n$, then
\begin{equation}
 \left|\E[\gen]\right|
 \leq\frac1n\sum_{i=1}^n
 \sqrt{2\sigma^2\hfun_{\epsilon_i}(P_{XY})}.
 \label{eq:individual-privacy}
\end{equation}
Equivalently, each summand is capped by
$\sqrt{2\sigma^2H(Y)}$ and by
$\sqrt{2\sigma^2[H(Y\mid X)+\epsilon_i]}$.
\end{theorem}

\begin{proof}
Applying the transfer lemma to $(W,X_i,Y_i)$ gives
 $I(W;Y_i)\leq\hfun_{\epsilon_i}(P_{XY}).$
Substitute this inequality into \eqref{eq:privacy-ismi}.
\end{proof}

\begin{corollary}[Block leakage implies an averaged individual bound]
\label{cor:block-to-individual}
Under \eqref{eq:selective-privacy}, independence of $X_1,\ldots,X_n$ gives
\begin{equation}
 \sum_{i=1}^n I(W;X_i)\leq I(W;X^n)\leq\epsilon_n.    \label{eq:block-sum}
\end{equation}
Consequently,
 $\left|\E[\gen]\right|
 \leq\sqrt{2\sigma^2
 \hfun_{\epsilon_n/n}(P_{XY})},$              
which agrees with \eqref{eq:privacy-main-explicit}.
\end{corollary}

\begin{proof}
Independence yields
\begin{align*}
 I(W;X^n)
 =H(X^n)-H(X^n\mid W)
 \geq\sum_{i=1}^nH(X_i)-\sum_{i=1}^nH(X_i\mid W)
 =\sum_{i=1}^nI(W;X_i).
\end{align*}
The map $u\mapsto\sqrt{\hfun_u(P_{XY})}$ is concave and nondecreasing.
Jensen's inequality and \eqref{eq:block-sum} finish the proof.
\end{proof}

\subsection{Per-output total-variation privacy}

Average mutual information may hide outputs with atypically large leakage.
Consider the stronger condition
\begin{equation}
 \left\|P_{X_i\mid W=w}-P_X\right\|_1\leq\delta_i,
 \quad\text{for all $w$ in the support of $W$}.        \label{eq:tv-privacy}
\end{equation}
Let $p_{\min}:=\min_{x:P_X(x)>0}P_X(x)>0$.

\begin{lemma}[From total variation to mutual information]
\label{lem:tv-to-mi}
Under \eqref{eq:tv-privacy},
 $I(W;X_i)\leq\frac{\delta_i^2}{p_{\min}}.$             
\end{lemma}

\begin{proof}
For distributions $P,Q$ on the same finite support,
\[
 D(P\|Q)\leq\sum_x\frac{(P(x)-Q(x))^2}{Q(x)}
 \leq\frac{\|P-Q\|_1^2}{\min_xQ(x)}.
\]
Apply this conditionally for every $W=w$ and average over $W$.
\end{proof}

\begin{theorem}[Total-variation selective privacy]
\label{thm:tv-privacy}
If \eqref{eq:tv-privacy} holds, then
\begin{equation}
 \left|\E[\gen]\right|
 \leq\frac1n\sum_{i=1}^n
 \sqrt{2\sigma^2
 \hfun_{\delta_i^2/p_{\min}}(P_{XY})}.               \label{eq:tv-privacy-bound}
\end{equation}
For $\delta_i=\delta$, the sum reduces to a single term.
\end{theorem}

\section{Adaptive Selective-Privacy Learning}
\label{sec:privacy-adaptive}

The one-shot bounds apply to a released model after training.  For a model
that is repeatedly updated on the same sample, the transcript itself can
leak information even if only one checkpoint is eventually selected.  Let
$\Pi_k=(V_1,\ldots,V_k)$ be generated as in~\eqref{eq:transcript-model} with
$D_n=(X^n,Y^n)$, and suppose
\begin{align}
 I(V_k;X^n\mid\Pi_{k-1})&\leq
       \epsilon_{n,k}^{\rm priv},                    \label{eq:adaptive-privacy-ledger}\\
 I(V_k;Y^n\mid\Pi_{k-1})&\leq
       \rho_{n,k}^{\rm task}.                        \label{eq:adaptive-task-ledger}
\end{align}
Define the cumulative ledgers
\begin{equation}
 E_{n,K}:=\sum_{k=1}^K\epsilon_{n,k}^{\rm priv},
 \qquad R_{n,K}:=\sum_{k=1}^K\rho_{n,k}^{\rm task}. \label{eq:privacy-ledger-sums}
\end{equation}

\begin{theorem}[Adaptive selective-privacy bound]
\label{thm:adaptive-privacy}
Let $\widehat W$ be a terminal model, an adaptively selected checkpoint, or a
bounded-stopping output that is a post-processing of $\Pi_K$.  Under
\eqref{eq:adaptive-privacy-ledger},
\begin{equation}
 |\E[\gen(\widehat W,Y^n)]|
 \leq\sqrt{2\sigma^2\min\left\{
 H(Y),H(Y\mid X)+\frac{E_{n,K}}n\right\}}.          \label{eq:adaptive-privacy-bound}
\end{equation}
If~\eqref{eq:adaptive-task-ledger} also holds, then
\begin{equation}
 |\E[\gen(\widehat W,Y^n)]|
 \leq\sqrt{2\sigma^2\min\left\{
 H(Y),H(Y\mid X)+\frac{E_{n,K}}n,
 \frac{R_{n,K}}n\right\}}.                         \label{eq:adaptive-privacy-compressed}
\end{equation}
The task ledger is implied by the stronger conditional full-sample budgets
\begin{equation}
 I(V_k;X^n,Y^n\mid\Pi_{k-1})\leq c_{n,k},           \label{eq:adaptive-full-ledger}
\end{equation}
in which case $R_{n,K}$ in~\eqref{eq:adaptive-privacy-compressed} may be
replaced by $\sum_{k=1}^Kc_{n,k}$.
\end{theorem}

\begin{proof}
Apply Theorem~\ref{thm:adaptive-transfer} with $(A,B)=(X,Y)$.  Under
\eqref{eq:adaptive-full-ledger}, the conditional chain rule gives
$I(V_k;Y^n\mid\Pi_{k-1})\leq
I(V_k;X^n,Y^n\mid\Pi_{k-1})$, which supplies the last statement.
\end{proof}

The theorem makes the scaling requirement transparent.  When $Y=f(X)$, the
privacy route vanishes whenever $E_{n,K}=o(n)$.  Thus a constant
per-update leakage permits $K=o(n)$ updates, while a summable update schedule
can remain informative for arbitrarily many updates.  When $H(Y\mid X)>0$,
the same residual floor as in the one-shot theorem remains; the task ledger
can remove it when $R_{n,K}=o(n)$.

\begin{theorem}[Individual-sample adaptive privacy]
\label{thm:adaptive-individual-privacy}
Suppose that for every record $i$ and update $k$,
\begin{equation}
 I(V_k;X_i\mid\Pi_{k-1})\leq\epsilon_{i,k}.          \label{eq:adaptive-coordinate-ledger}
\end{equation}
For any transcript-selected $\widehat W$,
\begin{equation}
 |\E[\gen(\widehat W,Y^n)]|
 \leq\frac1n\sum_{i=1}^n
 \sqrt{2\sigma^2\hfun_{\sum_{k=1}^K\epsilon_{i,k}}(P_{XY})}. \label{eq:adaptive-coordinate-bound}
\end{equation}
\end{theorem}

\begin{proof}
The chain rule yields
$I(\Pi_K;X_i)\leq\sum_k\epsilon_{i,k}$, and data processing gives the same
upper bound for $I(\widehat W;X_i)$.  Apply Lemma~\ref{lem:transfer} coordinatewise
and substitute the result into the individual-sample inequality
\eqref{eq:privacy-ismi}.
\end{proof}

\begin{figure}[t]
\centering
\begin{tikzpicture}
\begin{axis}[
 width=0.76\linewidth,height=0.46\linewidth,
 xlabel={Number of adaptive updates $K$},ylabel={Upper bound},
 xmin=1,xmax=100,ymin=0,ymax=0.11,
 grid=both,legend style={font=\small,at={(0.02,0.98)},anchor=north west},
 tick label style={font=\small},label style={font=\small},
 domain=1:100,samples=150]
\addplot+[thick] {sqrt(0.5*min(ln(2),0.1*x/500))};
\addlegendentry{$\epsilon_k=0.1$}
\addplot+[thick,dashed] {sqrt(0.5*min(ln(2),5*(1-pow(0.9,x))/500))};
\addlegendentry{$\epsilon_k=0.5(0.9)^{k-1}$}
\end{axis}
\end{tikzpicture}
\caption{Adaptive selective-privacy certificate for $n=500$, bounded loss,
and $H(Y\mid X)=0$.  The constant per-update ledger grows linearly, whereas
the geometric ledger has a finite cumulative information cost.}
\label{fig:adaptive-ledger}
\end{figure}
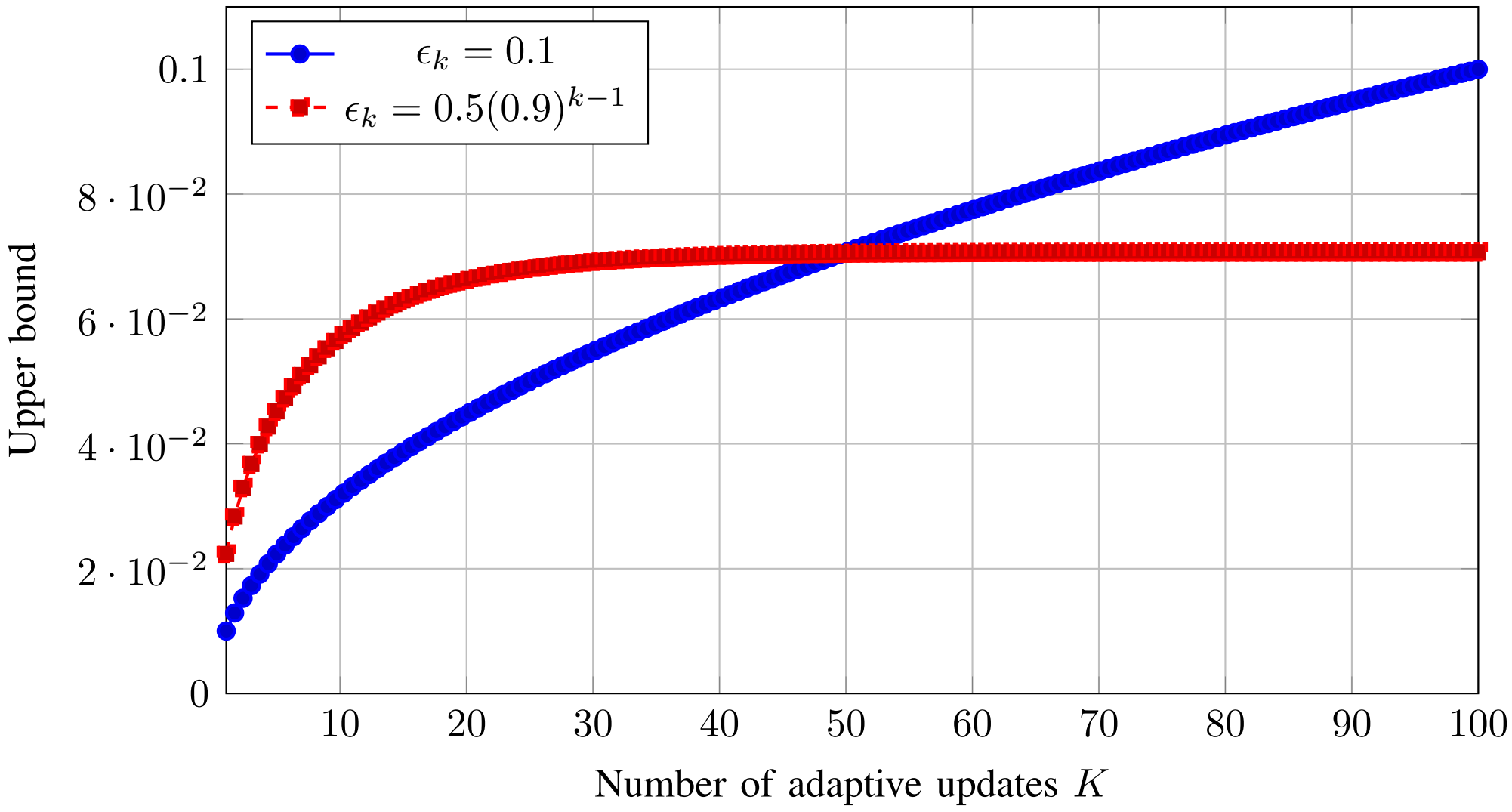

\section{Privacy Witnesses and Tightness}
\label{sec:privacy-witness}

The following results restore the functional-representation conclusions from
the selective-privacy manuscript without using a channel supremum to define
$\hfun_\epsilon$.  They are existence statements: at least one admissible
release can retain the stated amount of task information.  They do not lower-
bound the generalization error of every learner.

\begin{theorem}[EFRL and ESFRL witnesses]
\label{thm:privacy-witness}
Assume finite alphabets and
$0\leq\epsilon_n<nI(X;Y)$.  There exists an auxiliary output $U$, generated
from $(X^n,Y^n)$, such that $I(U;X^n)\leq\epsilon_n$ and
\begin{equation}
 I(U;Y^n)\geq
 \pos{nH(Y\mid X)-nH(X\mid Y)+\epsilon_n}.            \label{eq:efrl-witness}
\end{equation}
If $H(X)>0$, another construction satisfies the same leakage constraint and
\begin{align}
 I(U;Y^n)&\geq{}\Big[nH(Y\mid X)-\alpha_n nH(X\mid Y)+\epsilon_n
 -(1-\alpha_n)[\log(nI(X;Y)+3.4)+1]\Big]^+,      \label{eq:esfrl-witness}\\
 \alpha_n&:=\frac{\epsilon_n}{nH(X)}.                 \label{eq:alpha-privacy}
\end{align}
\end{theorem}

\begin{proof}[Proof outline]
Apply the extended functional representation and extended strong functional
representation constructions to the block pair $(X^n,Y^n)$
\cite{zamani2024privacy,zamani2026fairness,li2018strong,11686056}.  The i.i.d. identities
$H(Y^n\mid X^n)=nH(Y\mid X)$,
$H(X^n\mid Y^n)=nH(X\mid Y)$,
$I(X^n;Y^n)=nI(X;Y)$, and $H(X^n)=nH(X)$ yield the displayed expressions.
The positive part records nonnegativity of mutual information. A detailed
block substitution is given in Appendix~\ref{app:privacy-witness}.
\end{proof}

\begin{corollary}[Matching regime]
\label{cor:privacy-matching}
If $H(X\mid Y)=0$ and $0\leq\epsilon_n<nI(X;Y)$, the EFRL witness in
\eqref{eq:efrl-witness} attains
 $I(U;Y^n)=nH(Y\mid X)+\epsilon_n
 =\hfun_{\epsilon_n}^{(n)}(P_{XY}^{\otimes n}).$ 
Thus the direct information certificate is tight in this regime.
\end{corollary}

\begin{remark}[Common-information regimes]
Additional matching cases follow when the relevant common information equals
the mutual information \cite{zamani2024privacy}.  Such statements concern the
attainability of task information.  The generalization theorem itself remains
the uniform upper bound in Theorem~\ref{thm:privacy-main}.
\end{remark}

\section{Privacy Examples and Numerical Results}
\label{sec:privacy-examples}

\subsection{Analytical examples}

\begin{example}[Deterministic task]
Let $X\sim\mathrm{Bernoulli}(1/2)$ and $Y=X$.  Then $H(Y\mid X)=0$.
For $[0,1]$-bounded loss,
 $\left|\E[\gen]\right|
 \leq\sqrt{\frac{1}{2}\min\!\left\{\log2,
 \frac{\epsilon_n}{n}\right\}}.$               
For an $O(1)$ budget, this is $O(n^{-1/2})$.
\end{example}

\begin{example}[Noisy task]
\label{ex:privacy-bsc}
Let
\begin{equation}
 X\sim\mathrm{Bernoulli}(1/2),\qquad
 Y=X\oplus N,\qquad N\sim\mathrm{Bernoulli}(\theta),\label{eq:bsc-model}
\end{equation}
with $N\perp X$.  With binary entropy
 $h_2(\theta):=-\theta\log\theta-(1-\theta)\log(1-\theta),$
we have $H(Y)=\log2$ and $H(Y\mid X)=h_2(\theta)$. Hence
 $\left|\E[\gen]\right|
 \leq\sqrt{\frac12\min\!\left\{\log2,
 h_2(\theta)+\frac{\epsilon_n}{n}\right\}}.$  
For fixed $\theta>0$, increasing $n$ reduces the budget term but not the
residual-entropy term.
\end{example}

\begin{example}[Independent useful component]
Let $Y=(X,N)$ with $N\perp X$.  Then
$H(Y\mid X)=H(N)$.  Under perfect privacy, the certificate is
 $\left|\E[\gen]\right|
 \leq\sqrt{2\sigma^2\min\{H(Y),H(N)\}}.$       
The constraint on $X^n$ does not charge information carried only by $N^n$.
Task compression is therefore the direct way to remove this residual term.
\end{example}

\begin{example}[Privacy plus logarithmic compression]
In Example~\ref{ex:privacy-bsc}, suppose
$I(W;Y^n)\leq c\log n$.  Then
 $\left|\E[\gen]\right|
 \leq\sqrt{\frac1{2n}\min\!\left\{c\log n,
 n\log2,nh_2(\theta)+\epsilon_n\right\}},    $
which is $O(\sqrt{\log n/n})$.
\end{example}

\subsection{Bounds as functions of sample size and noise}

For the binary model \eqref{eq:bsc-model}, define, for $[0,1]$ loss,
 $B_{\mathrm{sel}}(n,\epsilon,\theta)
 :=\sqrt{\frac12\min\!\left\{\log2,
 h_2(\theta)+\frac{\epsilon}{n}\right\}},$  and            
 $B_{\mathrm{full}}(n,\epsilon)
 :=\sqrt{\frac{\epsilon}{2n}}.$              

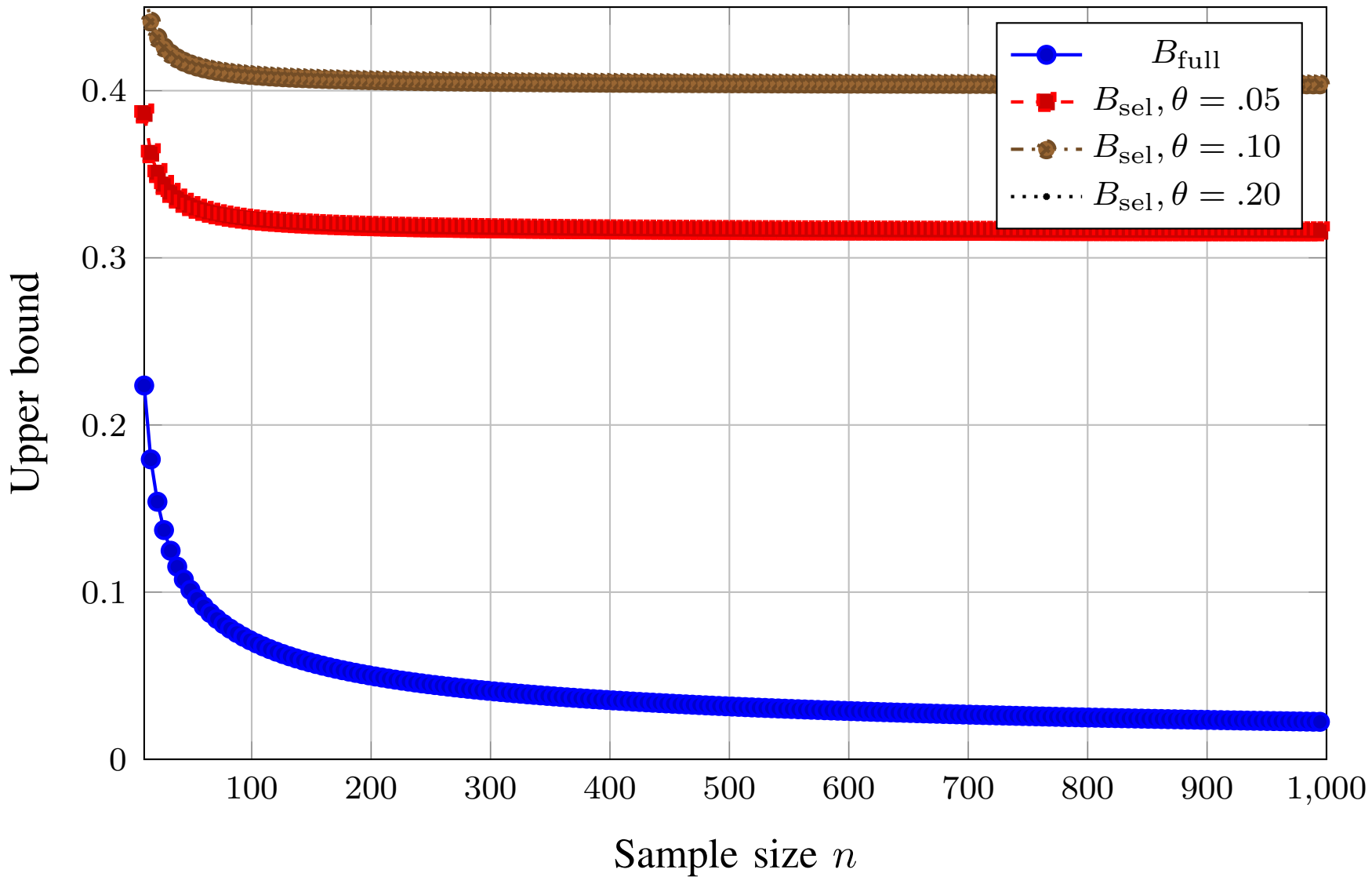
\begin{figure}[t]
\centering
\begin{tikzpicture}
\pgfmathdeclarefunction{binent}{1}{%
 \pgfmathparse{(#1==0 || #1==1) ? 0 : -(#1)*ln(#1)-(1-#1)*ln(1-#1)}%
}
\begin{axis}[
 width=0.70\linewidth,height=0.48\linewidth,
 xlabel={Sample size $n$},ylabel={Upper bound},
 xmin=10,xmax=1000,ymin=0,ymax=0.45,
 grid=both,legend style={font=\scriptsize,at={(0.98,0.98)},anchor=north east},
 tick label style={font=\scriptsize},label style={font=\small},
 domain=10:1000,samples=180]
\addplot+[thick] {sqrt(1/(2*x))};
\addlegendentry{$B_{\rm full}$}
\addplot+[thick,dashed] {sqrt(0.5*min(ln(2),binent(0.05)+1/x))};
\addlegendentry{$B_{\rm sel},\theta=.05$}
\addplot+[thick,dashdotted] {sqrt(0.5*min(ln(2),binent(0.10)+1/x))};
\addlegendentry{$B_{\rm sel},\theta=.10$}
\addplot+[thick,dotted] {sqrt(0.5*min(ln(2),binent(0.20)+1/x))};
\addlegendentry{$B_{\rm sel},\theta=.20$}
\end{axis}
\end{tikzpicture}
\caption{Selective-privacy and full-sample bounds for $\epsilon=1$.  Residual
noise creates a floor in the selective certificate.}
\label{fig:privacy-vs-n}
\end{figure}

Fig.~\ref{fig:privacy-vs-n} compares selective and full-sample information
control.  At $\theta=0$, the curves coincide in the active range.  For
$\theta>0$, the selective certificate approaches
$\sqrt{h_2(\theta)/2}$.

Define the compression-refined bound
 $B_{\mathrm{comp}}(n,\epsilon,\rho_n,\theta)
 :=\sqrt{\frac1{2n}\min\{\rho_n,n\log2,nh_2(\theta)+\epsilon\}}.$

\begin{figure}[t]
\centering
\begin{tikzpicture}
\pgfmathdeclarefunction{binent}{1}{%
 \pgfmathparse{(#1==0 || #1==1) ? 0 : -(#1)*ln(#1)-(1-#1)*ln(1-#1)}%
}
\begin{axis}[
 width=0.70\linewidth,height=0.48\linewidth,
 xlabel={Sample size $n$},ylabel={Upper bound},
 xmin=10,xmax=1000,ymin=0,ymax=0.45,
 grid=both,legend style={font=\scriptsize,at={(0.98,0.98)},anchor=north east},
 tick label style={font=\scriptsize},label style={font=\small},
 domain=10:1000,samples=180]
\addplot+[thick,dashed] {sqrt(0.5*min(ln(2),binent(0.10)+1/x))};
\addlegendentry{selective only}
\addplot+[thick] {sqrt((1/(2*x))*min(2*ln(x),min(x*ln(2),x*binent(0.10)+1)))};
\addlegendentry{$\rho_n=2\log n$}
\addplot+[thick,dashdotted] {sqrt((1/(2*x))*min(5*ln(x),min(x*ln(2),x*binent(0.10)+1)))};
\addlegendentry{$\rho_n=5\log n$}
\addplot+[thick,dotted] {sqrt(1/(2*x))};
\addlegendentry{full sample}
\end{axis}
\end{tikzpicture}
\caption{Compression-refined bounds for $\theta=0.10$ and $\epsilon=1$.
Sublinear task compression restores a vanishing certificate.}
\label{fig:privacy-compression}
\end{figure}
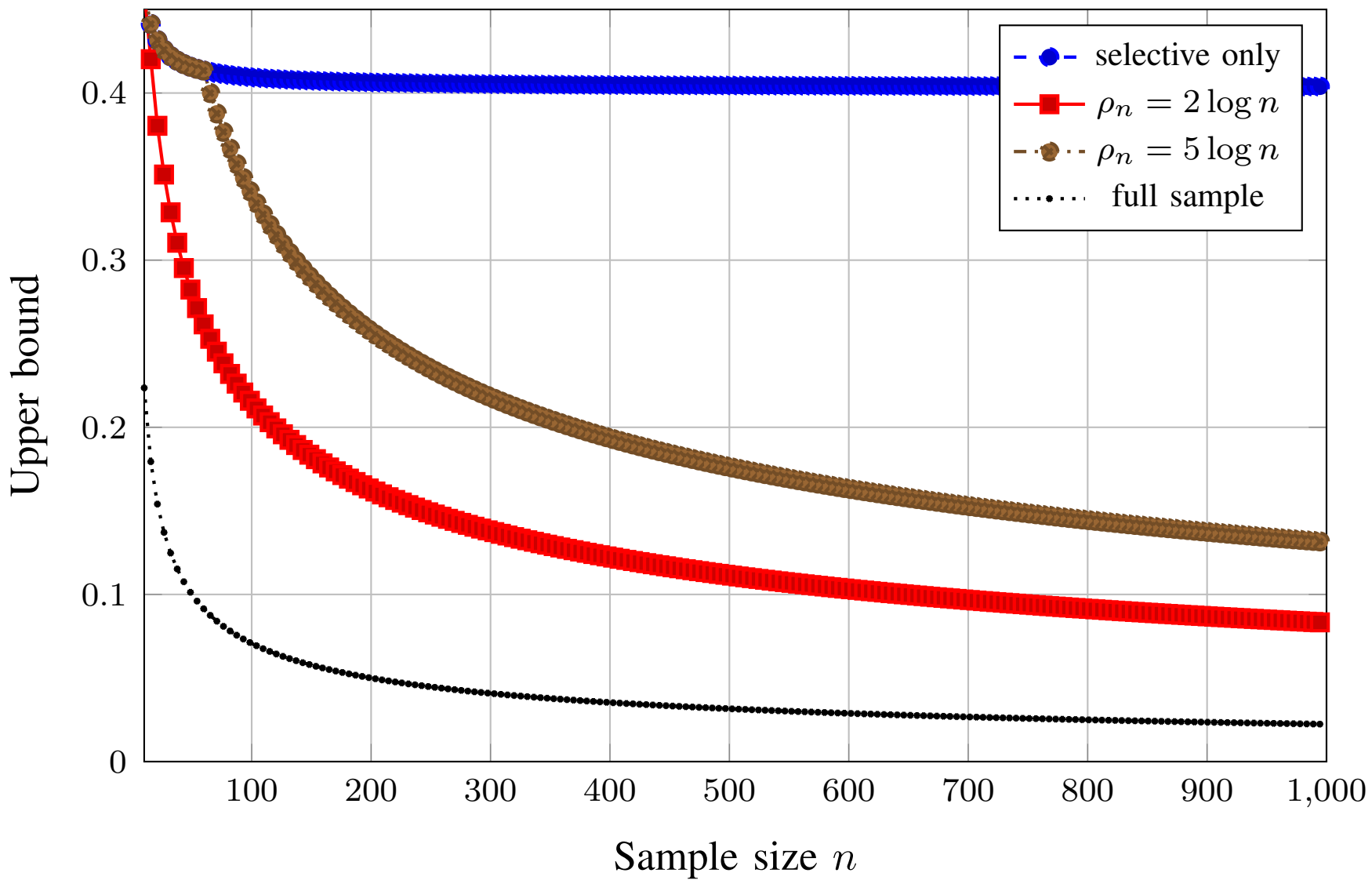

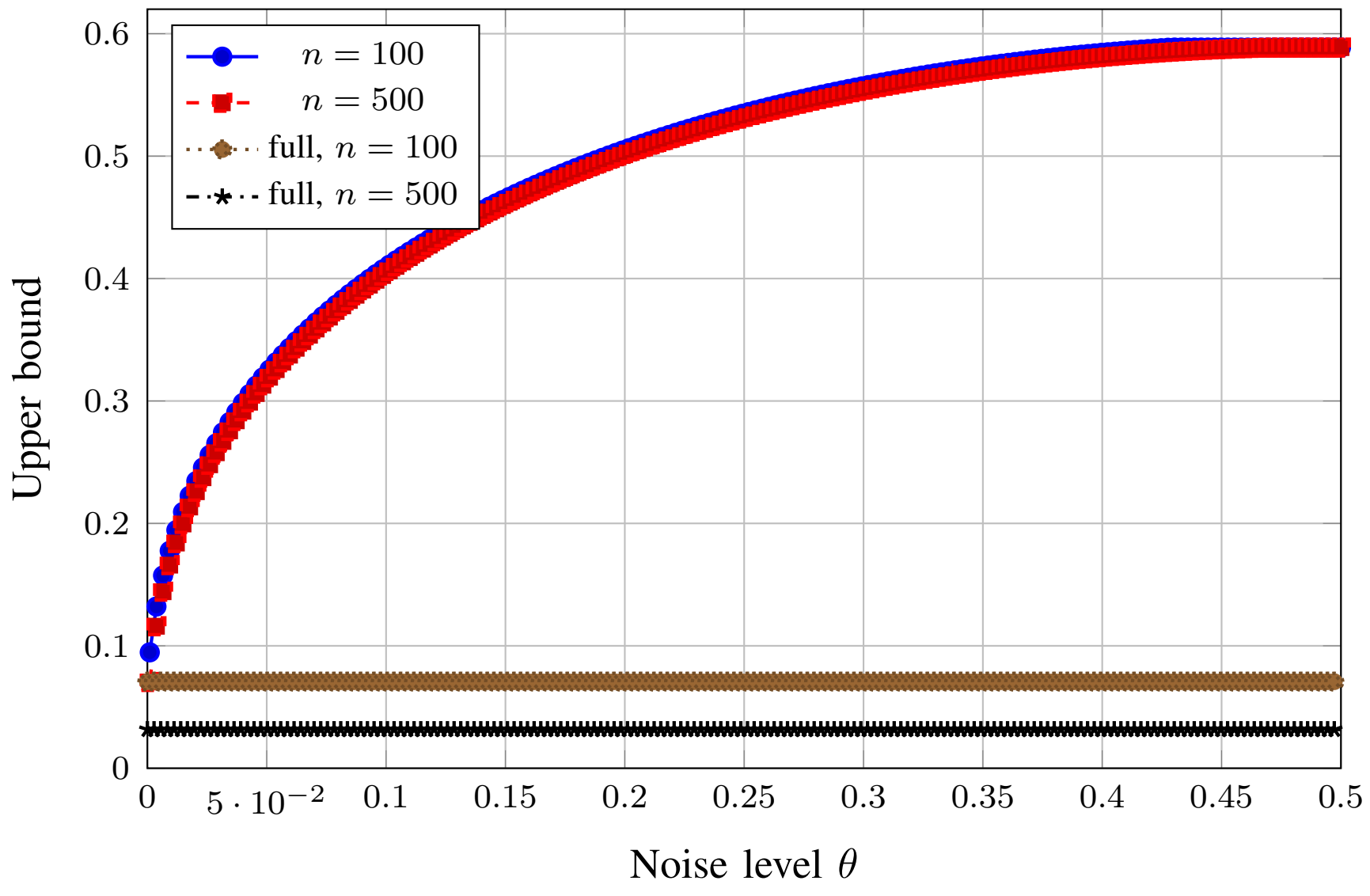
\begin{figure}[t]
\centering
\begin{tikzpicture}
\pgfmathdeclarefunction{binent}{1}{%
 \pgfmathparse{(#1==0 || #1==1) ? 0 : -(#1)*ln(#1)-(1-#1)*ln(1-#1)}%
}
\begin{axis}[
 width=0.70\linewidth,height=0.48\linewidth,
 xlabel={Noise level $\theta$},ylabel={Upper bound},
 xmin=0,xmax=0.5,ymin=0,ymax=0.62,
 grid=both,legend style={font=\scriptsize,at={(0.02,0.98)},anchor=north west},
 tick label style={font=\scriptsize},label style={font=\small},
 domain=0.001:0.5,samples=180]
\addplot+[thick] {sqrt(0.5*min(ln(2),binent(x)+1/100))};
\addlegendentry{$n=100$}
\addplot+[thick,dashed] {sqrt(0.5*min(ln(2),binent(x)+1/500))};
\addlegendentry{$n=500$}
\addplot+[thick,dotted,domain=0:0.5] {sqrt(1/(2*100))};
\addlegendentry{full, $n=100$}
\addplot+[thick,dashdotted,domain=0:0.5] {sqrt(1/(2*500))};
\addlegendentry{full, $n=500$}
\end{axis}
\end{tikzpicture}
\caption{Dependence of the selective-privacy certificate on residual task
noise for $\epsilon=1$.}
\label{fig:privacy-vs-theta}
\vspace{-2mm}
\end{figure}

Finally, Fig.~\ref{fig:privacy-vs-theta} holds $n$ fixed and varies task noise.
The entropy cap keeps the bound below $\sqrt{\log2/2}$.

\begin{table}[t]
\centering
\caption{Selective-privacy regimes for $[0,1]$-bounded loss.}
\label{tab:privacy-regimes}
\scriptsize
\begin{tabular}{@{}L{0.26\columnwidth}L{0.42\columnwidth}L{0.23\columnwidth}@{}}
\toprule
Model & Active information term per sample & Sufficient scaling \\
\midrule
$Y=f(X)$ & $\epsilon_n/n$ & $\epsilon_n=o(n)$ \\
$Y=X\oplus N$ & $h_2(\theta)+\epsilon_n/n$ & $\theta\to0$ and $\epsilon_n=o(n)$ \\
Task compression & $\rho_n/n$ & $\rho_n=o(n)$ \\
Adaptive transcript & $H(Y\mid X)+E_{n,K}/n$ & residual $\to0$ and $E_{n,K}=o(n)$ \\
\bottomrule
\end{tabular}
\end{table}

\section{Selective-Privacy Regime Analysis}
\label{sec:privacy-regimes}

The preceding results give several certificates for the same learner.  This
section makes their operating regimes explicit and separates fixed-source
asymptotics, budget saturation, and heterogeneous leakage.

\subsection{Fixed-source asymptotics and residual floors}

\begin{proposition}[Fixed-law limit]
\label{prop:privacy-limit}
Fix a finite-alphabet law $P_{XY}$. Suppose the normalized budget satisfies
$\epsilon_n/n\to e$, where $e\geq0$.  Then the right-hand side of
\eqref{eq:privacy-main-explicit} converges to
\begin{equation}
 \sqrt{2\sigma^2\min\{H(Y),H(Y\mid X)+e\}}.          \label{eq:privacy-limit}
\end{equation}
In particular, if $\epsilon_n=o(n)$, the limiting certificate is
 $\sqrt{2\sigma^2H(Y\mid X)}.$
It vanishes through the selective-privacy route for a fixed source precisely
when $H(Y\mid X)=0$.
\end{proposition}

\begin{proof}
The claim follows by continuity of the minimum and square-root maps.  Since
$H(Y\mid X)\leq H(Y)$, the second branch is active in the sublinear-budget
limit.  The statement concerns the certificate, not a necessary condition
for the learner's actual generalization error to vanish.
\end{proof}

The proposition explains the horizontal asymptotes in
Fig.~\ref{fig:privacy-vs-n}.  More data reduces the normalized leakage
$\epsilon_n/n$, but it does not change the source uncertainty left after the
private coordinate is revealed.  There are three ways around this floor.
First, the problem may be structurally deterministic.  Second, one may study
a sequence $P_{X_nY_n}$ with $H(Y_n\mid X_n)\to0$.  Third, an algorithmic
task-rate constraint can dominate the selective certificate.  The last route
is often the most practical because it does not require changing the data
distribution.

\subsection{Saturation of the privacy budget}

The entropy cap and the selective branch meet at
 $H(Y\mid X)+\frac{\epsilon_n}{n}=H(Y)
 \quad\Longleftrightarrow\quad
 \epsilon_n=nI(X;Y).$                         
Thus the direct certificate has two budget regimes.  For
$0\leq\epsilon_n<nI(X;Y)$, every additional nat of permitted private
information increases the task-information cap by at most one nat. At or
above $nI(X;Y)$, the trivial entropy cap $nH(Y)$ is active, and relaxing the
selective budget cannot further change the bound.  This is also why the
matching construction in Corollary~\ref{cor:privacy-matching} is stated below the
saturation threshold.

The threshold should not be interpreted as a recommended privacy level. It
only marks where the particular upper certificate stops responding to the
budget.  A learner can have $I(W;Y^n)$ far below the cap on either side of the
threshold.

\subsection{Global versus heterogeneous coordinate leakage}

The block and individual results answer different auditing questions.  A
single block budget is convenient when privacy is imposed on the complete
dataset.  Coordinate budgets $\epsilon_i$ are more informative when records
receive different protection levels or when the algorithm visits examples
unevenly.  Define
 $q(u):=\sqrt{2\sigma^2\hfun_u(P_{XY})}.$
Since $q$ is concave,
 $\frac1n\sum_{i=1}^n q(\epsilon_i)
 \leq q\left(\frac1n\sum_{i=1}^n\epsilon_i\right).$  
For a fixed total coordinate budget, the uniform allocation therefore gives
the largest value of this Jensen relaxation; strong heterogeneity can make
the individual bound smaller.  The result does not say that concentrating
leakage is desirable for privacy.  It says only that an averaged risk
certificate and a worst-record privacy criterion evaluate heterogeneity
differently.

Under the per-output total-variation condition, the effective coordinate
budget is $\delta_i^2/p_{\min}$.  If $H(Y\mid X)=0$ and all
$\delta_i=\delta_n$, Theorem~\ref{thm:tv-privacy} yields
 $|\E[\gen]|
 \leq \sqrt{\frac{2\sigma^2}{p_{\min}}}\,\delta_n,$  
until the entropy cap becomes active.  Hence vanishing per-output deviation
gives a vanishing certificate even without writing a block mutual-information
budget explicitly.  The dependence on $p_{\min}$ records the cost of
controlling likelihood ratios near rare private symbols.

\section{Part II: Fairness-Constrained Learning}
\label{sec:fair-model}

We now turn from protection of a private feature to bounded demographic
parity.  The formal similarity is useful, but the interpretation is
different: the constrained variable is a sensitive attribute, the feature
variable provides a second information bottleneck, and fairness itself must
also generalize from the training sample to the population.

\subsection{Statistical model and risk}

Let
 $Z=(S,X,T)\sim P_{SXT},\
 Z^n\sim P_{SXT}^{\otimes n},$                         
where $S\in\cS$ is sensitive, $X\in\cX$ is an observed feature, and
$T\in\cT$ is the task variable.  A randomized learner is a channel
 $P_{W\mid S^nX^nT^n}.$                                
For a loss $\ell:\cW\times\cT\to\R$, define
 $L_P(W):=\E_{T\sim P_T}[\ell(W,T)],\
 L_n(W):=\frac1n\sum_{i=1}^n\ell(W,T_i),\
 \gen(W,T^n):=L_P(W)-L_n(W).$                         
We assume throughout this part that $\ell(w,T)$ is $\sigma^2$-sub-Gaussian
under $P_T$ for each fixed $w$.

\subsection{Output-level fairness and compression}

We use the information-leakage form of bounded demographic parity
\begin{equation}
 I(W;S^n)\leq\epsilon_n,                              \label{eq:fair-constraint}
\end{equation}
and, when available, the feature-compression condition
\begin{equation}
 I(W;X^n)\leq r_n.                                   \label{eq:feature-compression}
\end{equation}
Condition~\eqref{eq:fair-constraint} says that the learned output carries at
most $\epsilon_n$ nats about the sensitive sample.  It is an output-level
constraint, rather than a pointwise differential-privacy guarantee.  The
compression budget in~\eqref{eq:feature-compression} is not a fairness
constraint; it supplies an independent route for controlling retained task
information.

The risk question and the fairness-generalization question must be kept
separate.  For task risk, the relevant information is $I(W;T^n)$.  For
population demographic parity of predictions $A_W(X)$, the relevant object
is the discrepancy between empirical and population laws of $(A_W(X),S)$.
The next two groups of results address these questions in turn.

\section{Task Generalization Under Fairness and Compression}
\label{sec:fair-main}

\subsection{A direct two-constraint certificate}

In direct analogy with~\eqref{eq:h-def}, define
\begin{align}
 \hfair_{\epsilon,r}(P_{SXT})
 :=\min\{\hfun_\epsilon(P_{ST}),\hfun_r(P_{XT})\}
 =\min\{H(T),H(T\mid S)+\epsilon,H(T\mid X)+r\}.
          \label{eq:hfair-direct}
\end{align}
This is the closed-form certificate obtained by intersecting the
sensitive-variable and feature-compression routes.  Its block counterpart is
\begin{align}
 \hfair_{\epsilon_n,r_n}(P_{SXT}^{\otimes n})
 =\min\{nH(T),nH(T\mid S)+\epsilon_n,
         nH(T\mid X)+r_n\}.                          \label{eq:hfair-block}
\end{align}

\begin{theorem}[Fair task bound]
\label{thm:fair-main}
If~\eqref{eq:fair-constraint} and~\eqref{eq:feature-compression} hold, then
\begin{equation}
 I(W;T^n)
 \leq \hfair_{\epsilon_n,r_n}(P_{SXT}^{\otimes n}),   \label{eq:fair-info-main}
\end{equation}
and hence
\begin{equation}
 \left|\E[\gen(W,T^n)]\right|
 \leq \sqrt{2\sigma^2\min\left\{
  H(T),\ H(T\mid S)+\frac{\epsilon_n}{n},
  H(T\mid X)+\frac{r_n}{n}
 \right\}}.                            \label{eq:fair-main-bound}
\end{equation}
For $[0,1]$-valued loss, $2\sigma^2$ can be replaced by $1/2$.
\end{theorem}

\begin{proof}
Apply Lemma~\ref{lem:transfer} first to $(A,B)=(S,T)$ and then to
$(A,B)=(X,T)$:
\begin{align}
 I(W;T^n)\leq nH(T\mid S)+I(W;S^n),\
 I(W;T^n)\leq nH(T\mid X)+I(W;X^n).
\end{align}
{\setlength{\emergencystretch}{1em}
The entropy bound $I(W;T^n)\leq nH(T)$ supplies the third branch.  Taking the
minimum proves~\eqref{eq:fair-info-main}; Lemma~\ref{lem:mi-generalization} then proves the
generalization statement.\par}
\end{proof}

The two conditional entropies have complementary meanings.  The term
$H(T\mid S)$ is task randomness not explained by the sensitive attribute,
and therefore not restricted by the fairness budget.  The term $H(T\mid X)$
is task randomness not explained by the feature, and therefore not restricted
by the feature-compression budget.  The smaller route is automatically
selected by~\eqref{eq:fair-main-bound}.

\begin{corollary}[Fairness-only route]
\label{cor:fairness-only}
Under~\eqref{eq:fair-constraint}, without assuming feature compression,
 $|\E[\gen]|
 \leq\sqrt{2\sigma^2
 \min\left\{H(T),H(T\mid S)+\frac{\epsilon_n}{n}\right\}}.$
If $H(T\mid S)=0$ and $\epsilon_n=O(1)$, this is $O(n^{-1/2})$.
\end{corollary}

\begin{corollary}[Compression-only route]
\label{cor:feature-only}
Under~\eqref{eq:feature-compression}, without assuming fairness,
 $|\E[\gen]|
 \leq\sqrt{2\sigma^2
 \min\left\{H(T),H(T\mid X)+\frac{r_n}{n}\right\}}.$
If $H(T\mid X)=0$ and $r_n=o(n)$, the certificate vanishes.
\end{corollary}

\begin{remark}[Fairness alone need not generalize]
Let $T=(S,N)$ with $N\perp S$.  A perfectly fair learner can retain $N^n$
while satisfying $I(W;S^n)=0$.  Since $H(T\mid S)=H(N)$, the fairness-only
upper bound has the nonvanishing floor $\sqrt{2\sigma^2H(N)}$.  This is the
fairness analogue of the residual task randomness in selective privacy.
\end{remark}

\subsection{Task-compression refinement}

Suppose the algorithm itself satisfies either
\begin{equation}
 I(W;T^n)\leq\rho_n                                 \label{eq:task-rate}
\end{equation}
or the stronger full-input constraint
$I(W;S^n,X^n,T^n)\leq\rho_n$.  This covers finite codebooks,
information-bottleneck training, quantized model outputs, and explicit
randomized compression.

\begin{theorem}[Three-way refinement]
\label{thm:fair-compression}
Under~\eqref{eq:fair-constraint},~\eqref{eq:feature-compression}, and
\eqref{eq:task-rate},
\begin{equation}
 |\E[\gen]|
 \leq \sqrt{\frac{2\sigma^2}{n}\min\left\{
  \rho_n,nH(T),nH(T\mid S)+\epsilon_n,
  nH(T\mid X)+r_n
 \right\}}.                            \label{eq:fair-threeway}
\end{equation}
Consequently, $\rho_n=O(\log n)$ gives
$|\E[\gen]|=O(\sqrt{\log n/n})$ independently of the residual entropies.
\end{theorem}

\begin{proof}
Intersect the task-rate bound with~\eqref{eq:fair-info-main}, then apply
Lemma~\ref{lem:mi-generalization}.
\end{proof}

The theorem identifies three qualitatively distinct regularizers: fairness
limits the sensitive route, feature compression limits the representation
route, and task compression limits direct task memorization.  A vanishing
bound requires only one of the three normalized information caps to vanish.

\section{Adaptive Fairness-Constrained Learning}
\label{sec:fair-adaptive}

Consider an adaptive transcript generated from
$D_n=(S^n,X^n,T^n)$.  At update $k$, impose the conditional ledgers
\begin{align}
 I(V_k;S^n\mid\Pi_{k-1})&\leq\epsilon_{n,k}^{\rm fair},
                                                        \label{eq:adaptive-fair-ledger}\\
 I(V_k;X^n\mid\Pi_{k-1})&\leq r_{n,k}^{\rm feat},
                                                        \label{eq:adaptive-feature-ledger}\\
 I(V_k;T^n\mid\Pi_{k-1})&\leq\rho_{n,k}^{\rm task}.
                                                        \label{eq:adaptive-fair-task-ledger}
\end{align}
Let
\begin{equation}
 E_{n,K}^{\rm fair}:=\sum_{k=1}^K\epsilon_{n,k}^{\rm fair},\quad
 R_{n,K}^{\rm feat}:=\sum_{k=1}^K r_{n,k}^{\rm feat},\quad
 C_{n,K}^{\rm task}:=\sum_{k=1}^K\rho_{n,k}^{\rm task}.       \label{eq:adaptive-fair-sums}
\end{equation}
The three sums measure different resources: sensitive dependence, feature
rate, and direct task memorization.

\begin{theorem}[Adaptive fairness and compression]
\label{thm:adaptive-fairness}
Let $\widehat W$ be any terminal model, transcript-selected checkpoint, or
bounded-stopping output.  Under
\eqref{eq:adaptive-fair-ledger} and~\eqref{eq:adaptive-feature-ledger},
\begin{equation}
 |\E[\gen(\widehat W,T^n)]|
 \leq\sqrt{2\sigma^2\min\left\{
  H(T),\ H(T\mid S)+\frac{E_{n,K}^{\rm fair}}n,
  H(T\mid X)+\frac{R_{n,K}^{\rm feat}}n
 \right\}}.                            \label{eq:adaptive-fair-bound}
\end{equation}
If~\eqref{eq:adaptive-fair-task-ledger} also holds, add
$C_{n,K}^{\rm task}/n$ as a fourth branch in the minimum.
\end{theorem}

\begin{proof}
The chain rule bounds $I(\Pi_K;S^n)$ by $E_{n,K}^{\rm fair}$ and
$I(\Pi_K;X^n)$ by $R_{n,K}^{\rm feat}$.  Apply
Theorem~\ref{thm:adaptive-transfer} to $(A,B)=(S,T)$ and $(A,B)=(X,T)$, intersect
the two certificates, and use data processing for $\widehat W$.  The task
ledger supplies the additional branch in the same way.
\end{proof}

The theorem remains meaningful when the number of updates is chosen from the
history.  What matters is the information spent before stopping, not the
nominal maximum number of updates.  Fairness of each checkpoint, however,
does not follow from a small sum unless the conditional sensitive-information
cost of every released part of the transcript is included.  In particular,
publishing validation scores or group-conditioned diagnostics without
charging them can invalidate the ledger.

\begin{corollary}[Heterogeneous fair-learning stages]
\label{cor:heterogeneous-fairness}
Let independent stage $j$ contain $n_j$ samples from
$P_{S_jX_jT_j}$, let $N=\sum_jn_j$, and suppose the stage-$j$ loss is
$\sigma_j^2$-sub-Gaussian.  Define
\begin{align}
 \mathsf H_T&:=\sum_jn_jH(T_j),&
 \mathsf H_{T\mid S}&:=\sum_jn_jH(T_j\mid S_j),\nonumber\\
 \mathsf H_{T\mid X}&:=\sum_jn_jH(T_j\mid X_j),&
 \mathsf V_T&:=\sum_jn_j\sigma_j^2.                 \label{eq:heterogeneous-fair-sums}
\end{align}
If the cumulative transcript ledgers for the aggregate sensitive and feature
blocks are $E_{N,K}^{\rm fair}$ and $R_{N,K}^{\rm feat}$, respectively, then
\begin{equation}
 |\E[\gen_{\rm het}(\widehat W)]|
 \leq\frac1N\sqrt{2\mathsf V_T\min\left\{
 \mathsf H_T,
 \mathsf H_{T\mid S}+E_{N,K}^{\rm fair},
 \mathsf H_{T\mid X}+R_{N,K}^{\rm feat}
 \right\}}.                                         \label{eq:heterogeneous-fair-bound}
\end{equation}
\end{corollary}

\begin{proof}
Apply Theorem~\ref{thm:heterogeneous-adaptive} once with protected coordinate $S$
and once with protected coordinate $X$, then take the minimum of the two
valid information caps.
\end{proof}

\section{Fairness Construction Witnesses and Tightness}
\label{sec:fair-witness}

The upper bounds above are universal: every learner obeying the stated
constraints satisfies them.  We next record all of the functional-
representation construction regimes associated with the fairness model.
They show that the conditional-entropy terms can be attained or approached by
some admissible representations.  Consistent with the paper's upper-bound
formulation, these constructions are not used to define
$\hfair_{\epsilon,r}$, and no supremum is introduced.
For compactness write
 $\cU_n(\epsilon_n,r_n):=
 \left\{U:I(U;S^n)\leq\epsilon_n,
             I(U;X^n)\leq r_n\right\}.$        

\begin{theorem}[Construction witnesses]
\label{thm:fair-witness}
Let $(S_i,X_i,T_i)$ be i.i.d. on finite alphabets.  In every regime there
exists a randomized variable $U$ generated from $(S^n,X^n,T^n)$ satisfying
the indicated constraints and the corresponding task-information guarantee.

For all $\epsilon_n,r_n\geq0$, there exists
$U_0\in\cU_n(\epsilon_n,r_n)$ such that
\begin{equation}
 I(U_0;T^n)\geq
 \pos{nH(T\mid X,S)-\log(nI(X,S;T)+3.4)-1}.          \label{eq:fair-L0}
\end{equation}
If
\begin{equation}
 0\leq r_n\leq nH(X\mid S)+\epsilon_n,              \label{eq:fair-regime-lowr}
\end{equation}
then there exists $U_1\in\cU_n(\epsilon_n,r_n)$ with
\begin{equation}
 I(U_1;T^n)\geq
 \pos{nH(T\mid X,S)+r_n-nH(X,S\mid T)}.             \label{eq:fair-L1}
\end{equation}
In the same regime, let
\begin{equation}
 \alpha_n:=\frac{r_n}{nH(X\mid S)+\epsilon_n}.       \label{eq:fair-alpha}
\end{equation}
Define also
\begin{align}
 \Lambda_n:={}(1-\alpha_n)nI(X,S;T)+\alpha_n n\min\{H(T),H(X,S)\}+3.4.               \label{eq:fair-lambda}
\end{align}
There exists an ESFRL-type witness $U_2\in\cU_n(\epsilon_n,r_n)$ satisfying
\begin{align}
 I(U_2;T^n)\geq{}\Big[nH(T\mid X,S)+r_n-\alpha_n nH(X,S\mid T)-1-\log\Lambda_n\Big]^+.
    \label{eq:fair-L2}
\end{align}
If
\begin{equation}
 nH(X\mid S)+\epsilon_n\leq r_n<nH(X),              \label{eq:fair-regime-highr}
\end{equation}
then some admissible $U_3$ satisfies
\begin{equation}
 I(U_3;T^n)\geq
 \pos{nH(T\mid S)-nH(S\mid T)+\epsilon_n}.          \label{eq:fair-L3}
\end{equation}
If
\begin{equation}
 r_n\leq\epsilon_n\leq nH(S\mid X)+r_n,            \label{eq:fair-regime-mideps}
\end{equation}
then some admissible $U_4$ satisfies
\begin{equation}
 I(U_4;T^n)\geq
 \pos{nH(T\mid X,S)+\epsilon_n-nH(X,S\mid T)}.      \label{eq:fair-L4}
\end{equation}
Finally, if
\begin{equation}
 nH(S\mid X)+r_n\leq\epsilon_n<nH(S),               \label{eq:fair-regime-higheps}
\end{equation}
then some admissible $U_5$ satisfies
\begin{equation}
 I(U_5;T^n)\geq
 \pos{nH(T\mid X)-nH(X\mid T)+r_n}.                 \label{eq:fair-L5}
\end{equation}
\end{theorem}

\begin{proof}[Proof construction]
Apply the corresponding one-shot constructions based on functional
representation to
the block variables
 $(S,X,T)\leftarrow(S^n,X^n,T^n).$
Memorylessness yields, for example,
\begin{align}
 H(T^n\mid S^n)&=nH(T\mid S),&
 H(T^n\mid X^n)&=nH(T\mid X),\nonumber\\
 H(T^n\mid X^n,S^n)&=nH(T\mid X,S),&
 I(X^n,S^n;T^n)&=nI(X,S;T),                         \label{eq:block-additivity}
\end{align}
and likewise for every entropy appearing in
\eqref{eq:fair-L1}--\eqref{eq:fair-L5}.  Substitution produces the displayed
block formulas. A more detailed accounting is given in
Appendix~\ref{app:witnesses}.
\end{proof}

\begin{corollary}[Matching deterministic regimes]
\label{cor:fair-tight}
If $S=f(T)$, then $H(S\mid T)=0$, so in the high-feature-rate regime the
witness in~\eqref{eq:fair-L3} meets the fairness branch
$nH(T\mid S)+\epsilon_n$ of~\eqref{eq:hfair-block}.  If $X=g(T)$, then
$H(X\mid T)=0$, so the witness in~\eqref{eq:fair-L5} meets the compression
branch $nH(T\mid X)+r_n$ in its stated regime.  Hence neither residual
entropy can be removed from a distribution-free upper bound based only on the
two information budgets.
\end{corollary}

\begin{remark}[What the witnesses do not claim]
The variables $U_j$ establish existence of information-rich admissible
representations.  They do not lower-bound the generalization error of every
fair learner, nor do they turn the upper-bound function into an extremal
definition.  A learner may discard far more task information than any
witness, and its actual generalization error may be much smaller than the
certificate.
\end{remark}

\section{Generalization of Fairness Metrics}
\label{sec:fair-metrics}

The preceding sections concern generalization of task risk under an
output-level fairness constraint.  A second question is whether a fairness
condition measured on the training data transfers to the population.  This
requires a complexity condition on the data-dependent predictor; small
$I(W;S^n)$ alone does not prevent $W$ from overfitting the feature sample.

Let $S\in\{0,1\}$, let $A_W(X)\in[0,1]$ be a deterministic or randomized
score, and write
\begin{equation}
 p_s:=\Prob(S=s),\qquad p_{\min}:=\min\{p_0,p_1\}>0. \label{eq:group-probs}
\end{equation}
The population demographic-parity gap is
\begin{equation}
 \Delta_P(W):=
 \left|\E[A_W(X)\mid S=1]-\E[A_W(X)\mid S=0]\right|.\label{eq:dp-pop}
\end{equation}
Define the oracle-normalized empirical gap
\begin{align}
 \widehat\Delta_n(W):=\bigg|&\frac1n\sum_{i=1}^n
       \frac{\ind\{S_i=1\}}{p_1}A_W(X_i)
 -\frac1n\sum_{i=1}^n
       \frac{\ind\{S_i=0\}}{p_0}A_W(X_i)\bigg|.       \label{eq:dp-empirical}
\end{align}
The use of $p_s$ isolates the adaptive-generalization effect.  Replacing
$p_s$ by empirical group frequencies introduces the familiar additional
concentration term for random group counts.

\begin{theorem}[Empirical-to-population parity gap]
\label{thm:dp-gap}
If
\begin{equation}
 I(W;S^n,X^n)\leq\kappa_n,                            \label{eq:adaptive-budget}
\end{equation}
then
\begin{equation}
 \E\left|\Delta_P(W)-\widehat\Delta_n(W)\right|
 \leq\left(\frac1{p_0}+\frac1{p_1}\right)
 \sqrt{\frac{\kappa_n+\log2}{2n}}.                  \label{eq:dp-gap-bound}
\end{equation}
Consequently, if $\widehat\Delta_n(W)\leq\tau$ almost surely, then
\begin{equation}
 \E[\Delta_P(W)]\leq\tau+
 \left(\frac1{p_0}+\frac1{p_1}\right)
 \sqrt{\frac{\kappa_n+\log2}{2n}}.                  \label{eq:dp-pop-transfer}
\end{equation}
\end{theorem}

\begin{proof}
For $s\in\{0,1\}$ set
\begin{equation}
 g_s(W,S,X):=\frac{\ind\{S=s\}}{p_s}A_W(X).
\end{equation}
For fixed $W$, $g_s$ lies in $[0,1/p_s]$ and is therefore
$1/(4p_s^2)$-sub-Gaussian by Hoeffding's lemma. The absolute-deviation
version of Lemma~\ref{lem:mi-generalization}, proved in Appendix~\ref{app:mi-lemmas}, gives
\begin{align}
 \E\bigg|\E[g_s(W,S,X)\mid W]
 -\frac1n\sum_{i=1}^ng_s(W,S_i,X_i)\bigg|
 \leq\frac1{p_s}\sqrt{\frac{\kappa_n+\log2}{2n}}.   \label{eq:group-one}
\end{align}
Sum~\eqref{eq:group-one} over $s=0,1$ and apply
$||a-b|-|c-d||\leq|a-c|+|b-d|$.  The final assertion follows from
$\Delta_P\leq\widehat\Delta_n+|\Delta_P-\widehat\Delta_n|$.
\end{proof}

The factor $1/p_0+1/p_1$ makes the statistical price of rare groups
explicit.  Fairness transfer thus needs both limited adaptivity
($\kappa_n=o(n)$) and nonvanishing group mass.  The result is orthogonal to
Theorem~\ref{thm:fair-main}: $\epsilon_n$ governs the fairness of the output, whereas
$\kappa_n$ governs the reliability of the empirical fairness assessment.

\subsection{Mutual-information demographic parity}

Suppose now that $A_W(X)$ takes values in a finite alphabet $\cA$ and
$S\in\cS$.  For a fixed realized $W$, let $P_{W,AS}$ denote the population
law of $(A_W(X),S)$ and let $\widehat P_{W,AS}^{(n)}$ be its training-sample
empirical law.  Put
\begin{equation}
 m:=|\cA||\cS|,\qquad
 \eta_n:=\frac m2\sqrt{\frac{\kappa_n+\log2}{2n}}.   \label{eq:eta-def}
\end{equation}
For $0\leq\eta\leq1/2$, define the continuity modulus
\begin{align}
 \omega_{\cA,\cS}(\eta):={}
\eta\log(|\cA|-1)+\eta\log(|\cS|-1)+\eta\log(|\cA||\cS|-1)+3h_2(\eta),              \label{eq:omega-def}
\end{align}
with $0\log0=0$ and the concave envelope used outside any interval on which
the displayed expression is not concave.

\begin{theorem}[Information-parity transfer]
\label{thm:mi-parity}
Under~\eqref{eq:adaptive-budget},
\begin{equation}
 \E\left[\left\|\widehat P_{W,AS}^{(n)}-P_{W,AS}\right\|_1\right]
 \leq m\sqrt{\frac{\kappa_n+\log2}{2n}}.             \label{eq:l1-transfer}
\end{equation}
If $\eta_n\leq1/2$, then
\begin{equation}
 \E\left|I_{\widehat P_n}(A_W;S)-I_P(A_W;S)\right|
 \leq\omega_{\cA,\cS}(\eta_n).                     \label{eq:mi-parity-transfer}
\end{equation}
In particular, empirical information parity
$I_{\widehat P_n}(A_W;S)\leq\epsilon$ implies
\begin{equation}
 \E[I_P(A_W;S)]
 \leq\epsilon+\omega_{\cA,\cS}(\eta_n).             \label{eq:mi-pop-fairness}
\end{equation}
\end{theorem}

\begin{proof}
For every $(a,s)\in\cA\times\cS$, apply the absolute information bound to
the indicator
\begin{equation}
 \phi_{a,s}(W,S,X):=\ind\{A_W(X)=a,S=s\}.
\end{equation}
Summing the $m$ coordinate bounds proves~\eqref{eq:l1-transfer}. Total
variation is half the $L_1$ distance.  Since
$I(A;S)=H(A)+H(S)-H(A,S)$, apply the finite-alphabet entropy-continuity bound
to the $A$, $S$, and $(A,S)$ marginals.  Jensen's inequality for the concave
modulus proves~\eqref{eq:mi-parity-transfer}; the last assertion follows by
the triangle inequality. Details are collected in
Appendix~\ref{app:entropy-continuity}.
\end{proof}

\begin{corollary}[Adaptive transfer of fairness metrics]
\label{cor:adaptive-fairness-metrics}
Let $\Pi_K=(V_1,\ldots,V_K)$ be an adaptive training transcript and suppose
\begin{equation}
 I(V_k;S^n,X^n\mid\Pi_{k-1})\leq\kappa_{n,k},
 \qquad k=1,\ldots,K.                               \label{eq:adaptive-kappa-ledger}
\end{equation}
Put $\kappa_{n,[K]}:=\sum_{k=1}^K\kappa_{n,k}$.  For every predictor
$\widehat W$ selected from the transcript, the parity-gap bounds
\eqref{eq:dp-gap-bound}--\eqref{eq:dp-pop-transfer} and the
information-parity bounds
\eqref{eq:l1-transfer}--\eqref{eq:mi-pop-fairness} remain valid after
replacing $\kappa_n$ by $\kappa_{n,[K]}$.
\end{corollary}

\begin{proof}
The chain rule and data processing give, respectively,
 $I(\Pi_K;S^n,X^n)\leq\sum_k\kappa_{n,k},$ and $\
 I(\widehat W;S^n,X^n)\leq I(\Pi_K;S^n,X^n).$
Apply
Theorems~\ref{thm:dp-gap} and~\ref{thm:mi-parity} with this cumulative budget.
\end{proof}

\section{Fairness Examples and Numerical Results}
\label{sec:fair-examples}

The following examples evaluate the certificate~\eqref{eq:hfair-direct} for
the original discrete models.

\begin{example}[Sensitive attribute determines the task]
\label{ex:fair-deterministic}
Let $S\sim\mathrm{Bernoulli}(1/2)$ and $T=S$.  Then $H(T\mid S)=0$, and for
$[0,1]$-bounded loss
\begin{equation}
 |\E[\gen]|\leq\sqrt{\frac{\epsilon_n}{2n}}.         \label{eq:fair-ex1}
\end{equation}
Thus bounded demographic parity alone gives the standard $n^{-1/2}$ rate
when $\epsilon_n=O(1)$.
\end{example}

\begin{example}[Residual task component]
\label{ex:fair-residual}
Let $T=(S,N)$, where $N\sim\mathrm{Bernoulli}(\theta)$ is independent of
$S$.  Then $H(T\mid S)=h_2(\theta)$.  Even perfect information parity permits
memorization of $N^n$, and the fairness-only bound is
 $|\E[\gen]|\leq\sqrt{\frac{h_2(\theta)}2}.$          
If additionally $I(W;T^n)\leq\rho_n$, then
 $|\E[\gen]|\leq
 \sqrt{\frac1{2n}\min\{\rho_n,nh_2(\theta)\}},$      
which vanishes for $\rho_n=o(n)$.
\end{example}

\begin{example}[Binary cascade]
\label{ex:binary-cascade}
Let
 $S\sim\mathrm{Bernoulli}(1/2),\ X=S\oplus E,
 \ T=X\oplus N,$                                 
where $E\sim\mathrm{Bernoulli}(q)$ and
$N\sim\mathrm{Bernoulli}(\theta)$ are independent.  With
 $q\star\theta:=q(1-\theta)+(1-q)\theta,$
we have $H(T\mid X)=h_2(\theta)$ and
$H(T\mid S)=h_2(q\star\theta)$.  Hence, for bounded loss,
\begin{equation}
 |\E[\gen]|
 \leq\sqrt{\frac12\min\left\{
  h_2(q\star\theta)+\frac{\epsilon_n}{n},
  h_2(\theta)+\frac{r_n}{n}
 \right\}}.                            \label{eq:cascade-bound}
\end{equation}
When $q\star\theta$ is small, the fairness route is strong; when $\theta$ is
small, the compression route is strong.
\end{example}

\begin{example}[Ternary matching regime]
\label{ex:ternary-tight}
Let $T\in\{0,1,2\}$ have law $P_T=(0.6,0.25,0.15)$.  Set $S=f(T)$ with
$f(0)=0$ and $f(1)=f(2)=1$, and set $X=g(T)=T$.
Then $H(S\mid T)=H(X\mid T)=H(T\mid X)=0$.  In the corresponding matching
regime of~Corollary~\ref{cor:fair-tight}, the direct certificate and a construction
witness coincide at
 $\min\{nH(T\mid S)+\epsilon_n,r_n\}.$         
The bounded-loss generalization upper bound is therefore
 $|\E[\gen]|\leq
 \sqrt{\frac1{2n}\min\{nH(T\mid S)+\epsilon_n,r_n\}}.$
The equality concerns attainable task information, not the realized
generalization error of every learner.
\end{example}

\begin{example}[Balanced-group fairness transfer]
\label{ex:balanced-fairness}
If $p_0=p_1=1/2$, Theorem~\ref{thm:dp-gap} becomes
 $\E|\Delta_P(W)-\widehat\Delta_n(W)|
 \leq4\sqrt{\frac{\kappa_n+\log2}{2n}}.    $    
Thus $\kappa_n=c\log n$ gives the demographic-parity transfer rate
 $O\left(\sqrt{\frac{\log n}{n}}\right).$
\end{example}

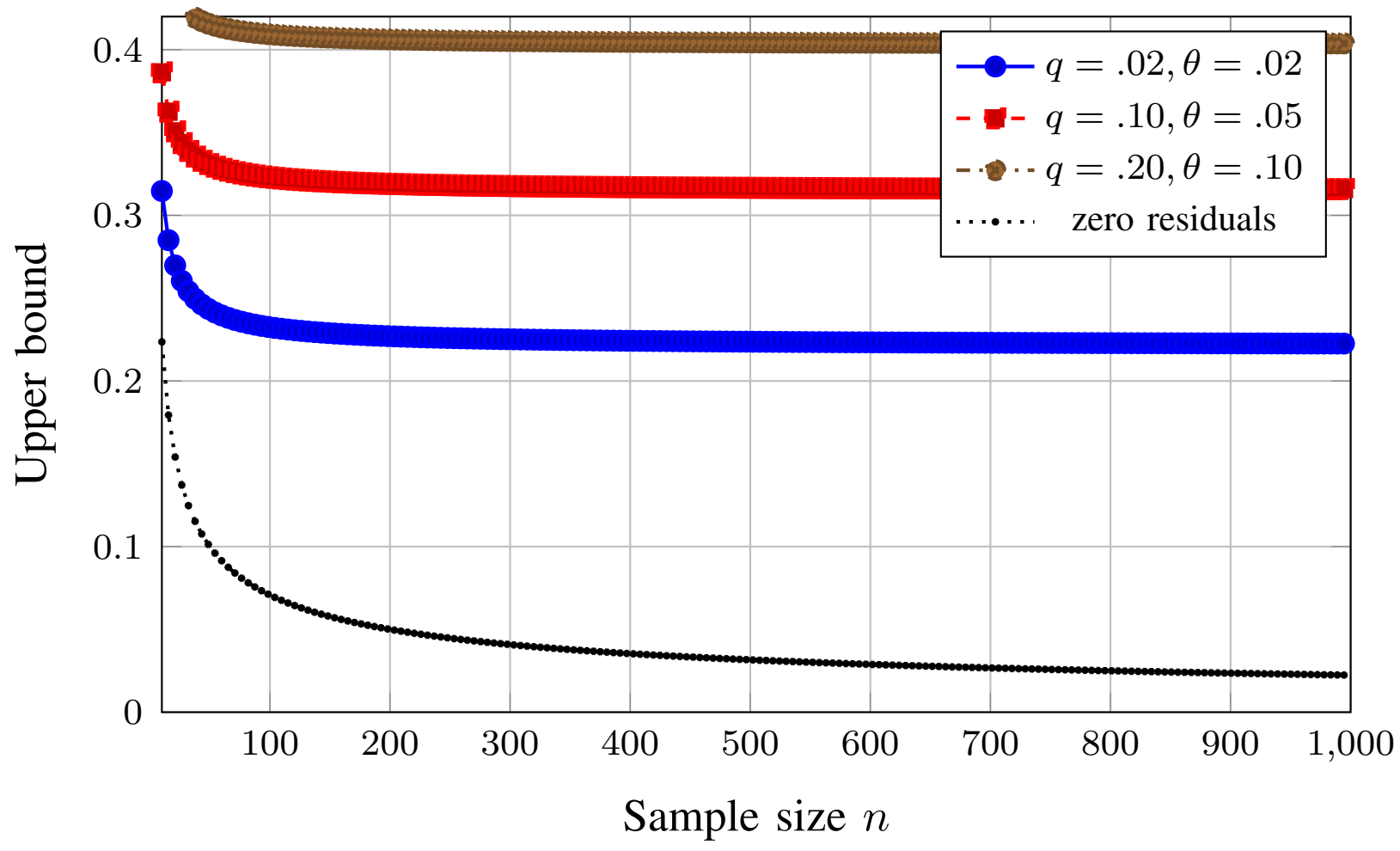
\begin{figure}[t]
\centering
\begin{tikzpicture}
\pgfmathdeclarefunction{binent}{1}{%
 \pgfmathparse{(#1==0 || #1==1) ? 0 : -(#1)*ln(#1)-(1-#1)*ln(1-#1)}%
}
\pgfmathdeclarefunction{bconv}{2}{%
 \pgfmathparse{#1*(1-#2)+(1-#1)*#2}%
}
\begin{axis}[
 width=0.65\linewidth,height=0.42\linewidth,
 xlabel={Sample size $n$},ylabel={Upper bound},
 xmin=10,xmax=1000,ymin=0,ymax=0.42,
 grid=both,legend style={font=\scriptsize,at={(0.98,0.98)},anchor=north east},
 tick label style={font=\scriptsize},label style={font=\small},
 domain=10:1000,samples=180]
\addplot+[thick] {sqrt(0.5*min(binent(bconv(0.02,0.02))+1/x,binent(0.02)+1/x))};
\addlegendentry{$q=.02,\theta=.02$}
\addplot+[thick,dashed] {sqrt(0.5*min(binent(bconv(0.10,0.05))+1/x,binent(0.05)+1/x))};
\addlegendentry{$q=.10,\theta=.05$}
\addplot+[thick,dashdotted] {sqrt(0.5*min(binent(bconv(0.20,0.10))+1/x,binent(0.10)+1/x))};
\addlegendentry{$q=.20,\theta=.10$}
\addplot+[thick,dotted] {sqrt(1/(2*x))};
\addlegendentry{zero residuals}
\end{axis}
\end{tikzpicture}
\caption{Fairness--compression bound for the binary cascade with
$\epsilon=r=1$.  Residual entropy creates a floor unless the task is nearly
determined by $S$ or $X$.}
\label{fig:cascade-vs-n}
\vspace{-2mm}
\end{figure}

For the binary cascade, set $\epsilon_n=\epsilon$, $r_n=r$, and define
\begin{equation}
 B(n,\epsilon,r,q,\theta):=
 \sqrt{\frac12\min\left\{
  h_2(q\star\theta)+\frac\epsilon n,
  h_2(\theta)+\frac rn
 \right\}}.                            \label{eq:cascade-B}
\end{equation}
Figure~\ref{fig:cascade-vs-n} 
illustrates numerical studies from the fairness model in Example \ref{ex:binary-cascade}.

\section{Fairness Regimes and the Two Generalization Problems}
\label{sec:fair-regimes}

Fair learning creates two distinct generalization problems.  The first asks
whether empirical task risk generalizes when the learned output is fair.  The
second asks whether a fairness metric evaluated on the training data
generalizes to the population.  The information budgets, residuals, and
failure modes are different, so it is useful to state their regimes together.

\subsection{Which task-information route is active?}
\label{sec:fair-route}

Ignoring the common entropy cap $H(T)$, the fairness branch of
\eqref{eq:fair-main-bound} is active exactly when
 $H(T\mid S)+\frac{\epsilon_n}{n}
 \leq H(T\mid X)+\frac{r_n}{n}.$              
Equivalently,
 $r_n-\epsilon_n
 \geq n\big[H(T\mid S)-H(T\mid X)\big].$      
The threshold depends on both source structure and budget imbalance.  Even if
$S$ predicts $T$ better than $X$ does, a loose fairness budget can make the
feature-compression route active.  Conversely, a tight fairness budget can
dominate a comparatively stringent feature budget when $S$ nearly determines
the task.

For the binary cascade of Example~\ref{ex:binary-cascade}, the route test becomes
 $h_2(q\star\theta)-h_2(\theta)
 \leq\frac{r_n-\epsilon_n}{n}.$                       
With equal fixed budgets, the compression branch is never worse because
$q\star\theta\geq\theta$ for $q,\theta\in[0,1/2]$ and binary entropy is
increasing on that interval.  The fairness route becomes competitive when
its budget is sufficiently tighter than the feature budget, or when the
source parameters lie outside that monotone comparison.  This analytic test
explains the lower envelope traced by the plots.

\begin{proposition}[Fixed-law fairness limit]
\label{prop:fair-limit}
For a fixed $P_{SXT}$, if $\epsilon_n/n\to e$ and $r_n/n\to r$, then the
certificate in Theorem~\ref{thm:fair-main} converges to
 $\sqrt{2\sigma^2\min\{H(T),H(T\mid S)+e,H(T\mid X)+r\}}.$
If both budgets are sublinear, the limiting residual is
 $\sqrt{2\sigma^2\min\{H(T\mid S),H(T\mid X)\}}.$     
\end{proposition}

The proof is immediate by continuity.  Importantly, both residuals need not
vanish: the minimum requires only one structural route.  If neither does,
sublinear task compression still yields a vanishing bound through
Theorem~\ref{thm:fair-compression}.

\subsection{Three budgets with three meanings}

The symbols $\epsilon_n$, $r_n$, and $\kappa_n$ should not be combined or
interchanged without an explicit implication:
 i. $\epsilon_n$ bounds information in $W$ about the sensitive sample and
 controls the fairness-to-task transfer.
 ii. $r_n$ bounds information in $W$ about the feature sample and controls
 the compression-to-task transfer.
 iii. $\kappa_n$ bounds adaptivity to $(S^n,X^n)$ and controls whether an
 empirical fairness statistic estimates its population counterpart.

A small $\epsilon_n$ does not generally imply a small $\kappa_n$: $W$ may
memorize $X^n$ while revealing almost nothing about $S^n$.  Likewise, a small
$r_n$ does not imply fairness if $W$ has direct access to $S^n$ or $T^n$.
Theorems~\ref{thm:fair-main}, \ref{thm:dp-gap}, and \ref{thm:mi-parity}
therefore answer complementary questions rather than progressively stronger
versions of one question.

\begin{table}[t]
\centering
\caption{Distinct fairness-related guarantees and their complexity inputs.}
\label{tab:fair-questions}
\scriptsize
\begin{tabular}{@{}L{0.28\columnwidth}L{0.29\columnwidth}L{0.34\columnwidth}@{}}
\toprule
Claim & Key budget & Uncontrolled term \\
\midrule
Task-risk transfer via fairness & $I(W;S^n)\leq\epsilon_n$ & $H(T\mid S)$ \\
Task-risk transfer via features & $I(W;X^n)\leq r_n$ & $H(T\mid X)$ \\
Parity-gap transfer & $I(W;S^n,X^n)\leq\kappa_n$ & rare-group factor \\
Information-parity transfer & same $\kappa_n$ & alphabet continuity modulus \\
Adaptive metric transfer & $\sum_k\kappa_{n,k}$ & cumulative transcript complexity \\
\bottomrule
\end{tabular}
\end{table}

\subsection{Sample complexity of population parity}

Suppose an empirical learner enforces $\widehat\Delta_n(W)\leq\tau$ and one
wants the additional population-transfer term in~\eqref{eq:dp-pop-transfer}
to be at most $\gamma>0$.  It is sufficient that
 $n\geq
 \frac{(\kappa_n+\log2)}{2\gamma^2}
 \left(\frac1{p_0}+\frac1{p_1}\right)^2.$      
Because $\kappa_n$ may itself depend on $n$, this inequality is implicit; for
$\kappa_n=O(\log n)$ it has the usual order
$\widetilde O(\gamma^{-2}p_{\min}^{-2})$.  The dependence on group mass is not
an artifact of information theory.  Each group-conditional mean uses an
inverse-probability weight, so rare groups have larger range and variance.
For mutual-information parity, the cellwise proof introduces the factor
$m=|\cA||\cS|$.  The resulting $L_1$ certificate is simple and fully
explicit, but it can be conservative for large output alphabets.  Coarsening
the prediction alphabet reduces $m$ while potentially changing the fairness
notion.  This is a genuine modeling tradeoff: the statistical cost of
certifying information parity grows with the resolution at which predictions
are audited.

\subsection{Reading the construction regimes}

The witness regimes in Theorem~\ref{thm:fair-witness} describe which information
route can be made rich under different relative budgets.  The low-$r_n$
regime exposes the joint unexplained task information $H(T\mid X,S)$ and the
penalty $H(X,S\mid T)$.  The high-$r_n$ regime isolates the fairness residual
$H(T\mid S)$, while the high-$\epsilon_n$ regime isolates the compression
residual $H(T\mid X)$.  Deterministic reverse maps remove the corresponding
penalty and produce matching information certificates.
These constructions are most useful as impossibility diagnostics.  If a
proposed universal upper bound deletes $H(T\mid S)$ or $H(T\mid X)$ while
using only the same budgets, a matching witness regime contradicts that
deletion.  They need not predict the behavior of a particular learning
algorithm, and they do not turn the risk upper bound into a two-sided risk
estimate.

\section{Unified View and Design Consequences }
\label{sec:discussion}

\subsection{One transfer principle, two constraint models}

The two parts of the paper are instances of the same information-transfer
argument.  A learner is constrained relative to a variable $A$, while its
loss is evaluated on a variable $B$.  The decomposition
 $I(W;B)=I(W;A)+I(W;B\mid A)-I(W;A\mid B)$     
turns a budget on $I(W;A)$ into a task-information certificate with residual
$H(B\mid A)$.  Privacy uses $(A,B)=(X,Y)$; fairness uses $(A,B)=(S,T)$,
while feature compression supplies the second pair $(A,B)=(X,T)$.

\begin{table*}[t]
\centering
\caption{Comparison of the privacy and fairness results.  All entries are
upper bounds for the particular learned output $W$.}
\label{tab:unified-results}
\scriptsize
\setlength{\tabcolsep}{3pt}
\begin{tabular}{@{}L{0.14\textwidth}L{0.18\textwidth}L{0.22\textwidth}L{0.22\textwidth}L{0.16\textwidth}@{}}
\toprule
Setting & Selective budget & Task-information certificate & Expected generalization & Additional control \\
\midrule
Privacy & $I(W;X^n)\leq\epsilon_n$ &
$\begin{aligned}\min\{&nH(Y),\\&nH(Y\mid X)+\epsilon_n\}\end{aligned}$ &
$\begin{aligned}&\big[2\sigma^2\min\{H(Y),\\&\quad H(Y\mid X)+\epsilon_n/n\}\big]^{1/2}\end{aligned}$ &
task rate $\rho_n$, individual leakage, TV privacy \\
Fairness & $I(W;S^n)\leq\epsilon_n$ &
$\begin{aligned}\min\{&nH(T),\\&nH(T\mid S)+\epsilon_n\}\end{aligned}$ &
$\begin{aligned}&\big[2\sigma^2\min\{H(T),\\&\quad H(T\mid S)+\epsilon_n/n\}\big]^{1/2}\end{aligned}$ &
feature budget $r_n$ \\
Fairness with compression & additionally $I(W;X^n)\leq r_n$ &
$\begin{aligned}\min\{&nH(T),\\&nH(T\mid S)+\epsilon_n,\\&nH(T\mid X)+r_n\}\end{aligned}$ &
square root of this minimum times $2\sigma^2/n$ &
task rate $\rho_n$ \\
Adaptive privacy & $I(V_k;X^n\mid\Pi_{k-1})\leq\epsilon_{n,k}$ &
$\min\{nH(Y),nH(Y\mid X)+\epsilon_{n,[K]}\}$ &
same privacy form with the cumulative ledger &
checkpoint selection and stopping \\
Adaptive fairness & conditional ledgers for $S^n$ and $X^n$ &
$\min\{nH(T),nH(T\mid S)+E_{n,K},nH(T\mid X)+R_{n,K}\}$ &
same fair form with cumulative ledgers &
heterogeneous independent stages \\
Fairness assessment & $I(W;S^n,X^n)\leq\kappa_n$ &
cellwise empirical-process control &
parity-gap and mutual-information transfer &
positive group mass, finite output alphabet \\
\bottomrule
\end{tabular}
\end{table*}
This view also explains why the privacy and fairness budgets cannot generally
replace the complete sample information in a standard generalization theorem.
The selective budget controls only one route to the task.  The residual
entropy is exactly the amount left unresolved after that route is observed.
The functional-representation witnesses show that this residual is not
merely a loose step in the chain-rule proof.

\subsection{When do the certificates vanish?}
For selective privacy, \eqref{eq:privacy-main-explicit} tends to zero under
the direct structural route
 $H(Y\mid X)\to0,\ \epsilon_n=o(n),$             
or under an independent task-compression route $\rho_n=o(n)$.  For fairness
with feature compression, it suffices that at least one of
 $H(T\mid S)+\epsilon_n/n\to0,
 H(T\mid X)+r_n/n\to0,
 \rho_n/n\to0$                                        
holds.  The minimum in~\eqref{eq:fair-threeway} makes these routes
substitutable from the standpoint of the certificate, even though they have
very different semantic and algorithmic meanings.
There is also an important order-of-limits issue.  With a fixed source law,
a strictly positive residual entropy remains a constant floor as $n$ grows.
A vanishing bound based on the selective constraint alone therefore requires
a sequence of problems for which the residual tends to zero, or an exactly
deterministic relation.  Sublinear compression, by contrast, supplies a
vanishing certificate without changing the source law.
For an adaptive learner, every one-shot budget in these conditions is
replaced by its cumulative conditional ledger.  Thus deterministic privacy
coupling gives a vanishing route when $E_{n,K}=o(n)$,
feature-determined fair learning gives one when
$R_{n,K}^{\rm feat}=o(n)$, and adaptive fairness-metric assessment requires
$\kappa_{n,[K]}=o(n)$.  These conditions permit the number of updates to
grow, but only at a rate compatible with the information spent per update.

\subsection{Implications for algorithm design}

The theory suggests a diagnostic sequence rather than a universal training
recipe.  First estimate which protected variable explains the task: this
determines whether the fairness or privacy route can be informative.  Next
measure the residual task entropy conditional on each constrained coordinate.
If all residuals are large, reducing a selective leakage budget alone cannot
certify generalization.  An explicit bottleneck on the model output or task
sample is then required.  Finally, if fairness is tuned empirically, control
$I(W;S^n,X^n)$ or a comparable adaptive-complexity quantity so that the
measured fairness constraint transfers to the population.
The distinction between $\epsilon_n$ and $\kappa_n$ is especially relevant.
A model can reveal little about the sensitive sample yet be highly adaptive
to the nonsensitive feature sample.  Such a model may satisfy output-level
information parity and still overfit its empirical parity score.  Conversely,
a stable model may generalize its measured parity gap even when that gap is
not small.  Constraint satisfaction and constraint generalization are
separate claims.
For sequential training, the same distinction becomes a bookkeeping rule.
The transcript should include every object used by a later decision:
checkpoints, group-conditioned validation scores, hyperparameter choices, and
stop indicators.  Conditional information budgets are accumulated in the
order in which those objects are revealed.  Auditing only the final model can
give a tighter number, but it does not certify a workflow in which earlier
releases were visible to an analyst or to downstream code.

\subsection{Robustness interpretation for adaptive deployment}

The bounds are robust to three procedural choices.  First, the kernels that
generate updates may depend arbitrarily on the previous transcript.  Second,
the reported checkpoint and stopping time may be selected from that
transcript.  Third, independent data batches may have heterogeneous source
laws and loss scales, as in Theorem~\ref{thm:heterogeneous-adaptive}.  In all three
cases the certificate depends on cumulative information and explicit
stagewise residuals, rather than on a prespecified update schedule.
This use of ``robust'' is deliberately delimited.  The results do not provide
worst-case guarantees under adversarial corruptions, dependent data streams,
or an unrestricted shift between the stage mixture and the deployment
population.  They give schedule-robust information accounting under the
stated product and independence assumptions.  Extending the selective
transfer principle to mixing processes, adversarial contamination, and
high-probability information-density control is an important direction for
future work.
\section{Conclusion}
\label{sec:conclusion}

This paper developed a unified information-theoretic formulation of
generalization for one-shot and adaptive learning under
selective privacy and fairness.
The function
 $\hfun_\epsilon(P_{AB})=\min\{H(B),H(B\mid A)+\epsilon\}$
is defined directly and used only as an upper-bound certificate.  In the
privacy model it yields block, compression-refined, individual-sample,
total-variation and sequential-update bounds.  In the
fairness model it combines with feature and task compression, composes across
adaptive updates, and is complemented by empirical-to-population guarantees
for both parity gaps and mutual-information parity.  A heterogeneous-stage
theorem covers independent nonstationary batches.  The full
FRL/EFRL/SFRL/ESFRL construction regimes are used to study the tightness of
the task-information certificates and to clarify where the residual entropies
are intrinsic.
The central message is that a selective constraint generalizes only the task
information statistically coupled to the constrained variable.  Residual
task randomness must be controlled by structure, compression, stability, or
another assumption.  Under repeated updates, the relevant budget is the sum
of conditional information increments, and a selected checkpoint is handled
as a post-processing of the charged transcript.  This separation makes
precise what privacy and fairness can certify by themselves, what they cannot
certify, and how additional information bottlenecks restore vanishing
guarantees for continually updated models.

\section*{Appendices}
These appendices contain the supplementary proofs, derivations, construction
substitutions, entropy calculations, and rate corollaries for the main
manuscript.
\appendices

\section{Mutual-Information Generalization Lemmas}
\label{app:mi-lemmas}

This appendix records the two versions of the information-theoretic
generalization inequality used throughout the paper.  Let $B^n\sim
P_B^{\otimes n}$ and let $W$ be jointly distributed with $B^n$.  For fixed
$w$, define
\begin{equation}
 F_w(B^n):=\frac1n\sum_{i=1}^n
 \big(\E_{B\sim P_B}\ell(w,B)-\ell(w,B_i)\big).
                                                               \label{eq:Fw}
\end{equation}
If $\ell(w,B)$ is $\sigma^2$-sub-Gaussian, then $F_w(B^n)$ is
$\sigma^2/n$-sub-Gaussian under the product law.

\begin{appendixlemma}[Decoupling inequality]
\label{lem:decoupling}
Let $(V,Q)$ have joint law $P_{VQ}$, and suppose $f(V,Q)$ is centered and
$\nu^2$-sub-Gaussian under the product law $P_VP_Q$.  Then
\begin{equation}
 |\E_{P_{VQ}}f(V,Q)|\leq\sqrt{2\nu^2I(V;Q)}.         \label{eq:decoupling}
\end{equation}
\end{appendixlemma}

\begin{proof}
For $\lambda>0$, the variational representation of relative entropy gives
\begin{align}
 \lambda\E_{P_{VQ}}f
 \leq D(P_{VQ}\|P_VP_Q)
 +\log\E_{P_VP_Q}e^{\lambda f}
 \leq I(V;Q)+\frac{\lambda^2\nu^2}{2}.             \label{eq:dv-step}
\end{align}
Divide by $\lambda$ and optimize at
$\lambda=\sqrt{2I(V;Q)/\nu^2}$.  Applying the same argument to $-f$ proves
the absolute value.
\end{proof}

Taking $(V,Q,f)=(W,B^n,F_W)$ proves
\begin{equation}
 |\E[\gen(W,B^n)]|
 \leq\sqrt{\frac{2\sigma^2}{n}I(W;B^n)},            \label{eq:app-mi-gen}
\end{equation}
which is the first inequality in Lemma~\ref{lem:mi-generalization}.

For fairness-metric transfer we need an expected absolute deviation.  Let
$G(W,B^n)$ be the difference between a population mean and its sample mean,
with range-$c$ summands.  Introduce an independent Rademacher sign
$R\in\{-1,1\}$ and choose it conditionally as the sign of $G$.  Then
$\E|G|=\E[RG]$ and
\begin{equation}
 I((W,R);B^n)\leq I(W;B^n)+H(R)\leq I(W;B^n)+\log2. \label{eq:sign-info}
\end{equation}
Hoeffding's lemma gives variance proxy $c^2/(4n)$, so
\begin{equation}
 \E|G(W,B^n)|
 \leq c\sqrt{\frac{I(W;B^n)+\log2}{2n}}.            \label{eq:absolute-mi}
\end{equation}
Equation~\eqref{eq:absolute-mi} yields~\eqref{eq:group-one}
with $c=1/p_s$
and the cellwise bounds in Theorem~\ref{thm:mi-parity}
with $c=1$.

\section{Adaptive and Heterogeneous-Stage Proofs}
\label{app:adaptive-proofs}

The sequential identity used throughout is exact.  For every $k$,
\begin{equation}
 I(\Pi_k;A^n)
 =\sum_{j=1}^k I(V_j;A^n\mid\Pi_{j-1}),             \label{eq:app-sequential-chain}
\end{equation}
and the same identity holds with $A^n$ replaced by any other data block.
This shows why update budgets must be conditional on the complete preceding
transcript. Unconditional bounds $I(V_j;A^n)\leq\epsilon_j$ do not in
general control the left-hand side of~\eqref{eq:app-sequential-chain}, because
different updates can reveal complementary information.

We prove Theorem~\ref{thm:heterogeneous-adaptive}.
Concatenate the independent
stage blocks as
\begin{equation}
 A^{[m]}=(A_1^{n_1},\ldots,A_m^{n_m}),\qquad
 B^{[m]}=(B_1^{n_1},\ldots,B_m^{n_m}).              \label{eq:app-aggregate}
\end{equation}
Independence within and across stages gives
\begin{align}
 H(B^{[m]})&=\sum_{j=1}^m n_jH(B_j)=\mathsf H_B,\nonumber\\
 H(B^{[m]}\mid A^{[m]})
 &=\sum_{j=1}^m n_jH(B_j\mid A_j)=\mathsf H_{B\mid A}. \label{eq:app-heterogeneous-entropy}
\end{align}
The conditional ledger and~\eqref{eq:app-sequential-chain} imply
\begin{equation}
 I(\Pi_K;A^{[m]})\leq\epsilon_{N,[K]}.              \label{eq:app-heterogeneous-ledger}
\end{equation}
The direct transfer identity therefore yields
\begin{equation}
 I(\widehat W;B^{[m]})
 \leq\min\{\mathsf H_B,
             \mathsf H_{B\mid A}+\epsilon_{N,[K]}\}. \label{eq:app-heterogeneous-info}
\end{equation}

For a fixed $w$, the $N$ centered summands in
$\gen_{\rm het}(w)$ are independent.  The log moment-generating function of
their $1/N$-weighted sum is at most
\begin{equation}
 \frac{\lambda^2}{2N^2}\sum_{j=1}^m n_j\sigma_j^2
 =\frac{\lambda^2\mathsf V}{2N^2}.                 \label{eq:app-heterogeneous-mgf}
\end{equation}
Apply Lemma~\ref{lem:decoupling} with variance proxy $\mathsf V/N^2$ and insert
\eqref{eq:app-heterogeneous-info} to obtain
\begin{equation*}
 |\E[\gen_{\rm het}(\widehat W)]|
 \leq\frac1N\sqrt{2\mathsf V
 \min\{\mathsf H_B,\mathsf H_{B\mid A}+\epsilon_{N,[K]}\}}.
\end{equation*}
If the task ledger is available, the chain
rule also gives $I(\Pi_K;B^{[m]})\leq\rho_{N,[K]}$, which adds the stated
branch.  The proof of Corollary~\ref{cor:heterogeneous-fairness}
repeats the transfer
step with $A^{[m]}=S^{[m]}$ and $A^{[m]}=X^{[m]}$ and intersects the two
valid task-information caps.

\section{Direct Transfer Identities}
\label{app:direct-transfer}

For completeness, we give the exact identity behind every direct
certificate.  For arbitrary jointly distributed $(U,A,B)$,
\begin{align}
 I(U;B)
 =H(B)-H(B\mid U)
 =I(U;A)+I(U;B\mid A)-I(U;A\mid B).                 \label{eq:transfer-exact}
\end{align}
Therefore
\begin{align}
 I(U;B)
 &\leq I(U;A)+I(U;B\mid A)
 \leq I(U;A)+H(B\mid A),                           \label{eq:transfer-direct-a}\\
 I(U;B)&\leq H(B).                                   \label{eq:transfer-direct-b}
\end{align}
If $I(U;A)\leq\epsilon$, the minimum of
\eqref{eq:transfer-direct-a} and~\eqref{eq:transfer-direct-b} is precisely
$\hfun_\epsilon(P_{AB})$.

For i.i.d. blocks, entropy additivity gives
\begin{align}
 H(B^n)&=nH(B),\nonumber\\
 H(B^n\mid A^n)&=nH(B\mid A),                       \label{eq:entropy-tensor}
\end{align}
and hence
\begin{equation}
 I(U;B^n)\leq\min\{nH(B),nH(B\mid A)+\epsilon_n\}. \label{eq:block-direct}
\end{equation}

The fairness result applies~\eqref{eq:block-direct} twice, with
$(A,B)=(S,T)$ and $(A,B)=(X,T)$, and intersects the two
certificates.  These substitutions prove all entropy upper bounds in the main
text.

\section{Details for the Selective-Privacy Refinements}
\label{app:privacy-details}

\subsection{From block leakage to individual leakage}

When $X_1,\ldots,X_n$ are independent,
\begin{align}
 I(W;X^n)
 &=H(X^n)-H(X^n\mid W)
 =\sum_{i=1}^nH(X_i)-H(X^n\mid W)
 \geq\sum_{i=1}^n\big[H(X_i)-H(X_i\mid W)\big]\nonumber\\
 &=\sum_{i=1}^nI(W;X_i),                             \label{eq:individual-sum-proof}
\end{align}
where conditional subadditivity gives the inequality.  Let
\begin{equation}
 q(u):=\sqrt{\min\{H(Y),H(Y\mid X)+u\}}.
\end{equation}
This function is nondecreasing and concave, and so
\begin{align}
 \frac1n\sum_{i=1}^nq(I(W;X_i))
 \leq q\left(\frac1n\sum_{i=1}^nI(W;X_i)\right)\leq q(\epsilon_n/n).                             \label{eq:jensen-individual}
\end{align}
Combining~\eqref{eq:jensen-individual} with the individual-sample
generalization inequality proves Corollary~\ref{cor:block-to-individual}.

\subsection{Total-variation conversion}

For probability mass functions $P,Q$ with $Q(x)>0$ on the support of $P$,
$\log t\leq t-1$ implies
\begin{align}
 D(P\|Q)
 \leq\sum_x\frac{(P(x)-Q(x))^2}{Q(x)}
 \leq\frac1{\min_xQ(x)}\left(\sum_x|P(x)-Q(x)|\right)^2.
\label{eq:kl-tv-detail}
\end{align}
With $P=P_{X_i\mid W=w}$ and $Q=P_X$, average
\eqref{eq:kl-tv-detail} over $w$ to obtain
$I(W;X_i)\leq\delta_i^2/p_{\min}$.  The individual selective-privacy result
then proves Theorem~\ref{thm:tv-privacy}.

\section{Block Construction Substitutions}
\label{app:witnesses}

This appendix spells out how the one-shot construction statements become
the block witnesses used for tightness diagnostics.  It does not introduce
an extremal definition of either certificate.

\subsection{Selective privacy}
\label{app:privacy-witness}

The EFRL construction for a generic finite pair $(A,B)$ supplies an output
$U$ meeting a leakage budget $\epsilon$ and a retained-information guarantee
of the form
\begin{equation}
 \pos{H(B\mid A)-H(A\mid B)+\epsilon}.               \label{eq:one-efrl-form}
\end{equation}
Set $(A,B)=(X^n,Y^n)$.  Product structure yields
\begin{equation}
 H(Y^n\mid X^n)-H(X^n\mid Y^n)+\epsilon_n
 =nH(Y\mid X)-nH(X\mid Y)+\epsilon_n,               \label{eq:block-efrl-sub}
\end{equation}
which is~\eqref{eq:efrl-witness}.

For the ESFRL construction the mixing coefficient is
\begin{equation}
 \alpha_n=\frac{\epsilon_n}{H(X^n)}
 =\frac{\epsilon_n}{nH(X)},                         \label{eq:block-alpha-sub}
\end{equation}
and the logarithmic term becomes
\begin{equation}
 \log(I(X^n;Y^n)+3.4)+1
 =\log(nI(X;Y)+3.4)+1.                                \label{eq:block-log-sub}
\end{equation}
Substitution gives~\eqref{eq:esfrl-witness}.
If $H(X\mid Y)=0$, the EFRL
expression equals $nH(Y\mid X)+\epsilon_n$, explaining
Corollary~\ref{cor:privacy-matching}.

\subsection{Fairness with feature compression}

For $(S^n,X^n,T^n)$, every entropy in a one-shot construction is additive:
\begin{align}
 H(T^n\mid X^n,S^n)&=nH(T\mid X,S),
 H(X^n,S^n\mid T^n)=nH(X,S\mid T),\nonumber\\
 H(X^n\mid S^n)&=nH(X\mid S),
 H(S^n\mid X^n)=nH(S\mid X),\nonumber\\
 H(S^n\mid T^n)&=nH(S\mid T),
 H(X^n\mid T^n)=nH(X\mid T).                       \label{eq:fair-all-additive}
\end{align}
The low-$r_n$ construction therefore produces
\begin{equation}
 \pos{nH(T\mid X,S)+r_n-nH(X,S\mid T)},             \label{eq:app-L1}
\end{equation}
while its ESFRL mixture uses
$\alpha_n=r_n/[nH(X\mid S)+\epsilon_n]$ and
\begin{align}
 &(1-\alpha_n)I(X^n,S^n;T^n)
 +\alpha_n\min\{H(T^n),H(X^n,S^n)\}+3.4\nonumber\\
 &\qquad=(1-\alpha_n)nI(X,S;T)
 +\alpha_n n\min\{H(T),H(X,S)\}+3.4.               \label{eq:app-L2-log}
\end{align}
The remaining regimes follow by the same substitution and yield,
respectively,
\begin{align}
 &\pos{nH(T\mid S)-nH(S\mid T)+\epsilon_n},\nonumber\\
 &\pos{nH(T\mid X,S)+\epsilon_n-nH(X,S\mid T)},\nonumber\\
 &\pos{nH(T\mid X)-nH(X\mid T)+r_n}.                \label{eq:app-L345}
\end{align}
The stated regime boundaries also tensorize because each entropy is additive.
This proves the complete list in Theorem~\ref{thm:fair-witness}.

\begin{table*}[t]
\centering
\caption{Construction witnesses in the fairness model.  Each row asserts
existence of an admissible $U$; none is a lower bound for every learner's
generalization error.}
\label{tab:witness-regimes}
\small
\begin{tabular}{@{}L{0.26\textwidth}L{0.31\textwidth}L{0.35\textwidth}@{}}
\toprule
Regime & Method/type & Guaranteed $I(U;T^n)$ \\
\midrule
All $\epsilon_n,r_n\geq0$ & strong functional representation &
$[nH(T\mid X,S)-\log(nI(X,S;T)+3.4)-1]^+$ \\
$0\leq r_n\leq nH(X\mid S)+\epsilon_n$ & EFRL &
$[nH(T\mid X,S)+r_n-nH(X,S\mid T)]^+$ \\
Same low-$r_n$ regime & ESFRL mixture & expression in~\eqref{eq:fair-L2}\\
$nH(X\mid S)+\epsilon_n\leq r_n<nH(X)$ & fairness-limited &
$[nH(T\mid S)-nH(S\mid T)+\epsilon_n]^+$ \\
$r_n\leq\epsilon_n\leq nH(S\mid X)+r_n$ & leakage-limited &
$[nH(T\mid X,S)+\epsilon_n-nH(X,S\mid T)]^+$ \\
$nH(S\mid X)+r_n\leq\epsilon_n<nH(S)$ & compression-limited &
$[nH(T\mid X)-nH(X\mid T)+r_n]^+$ \\
\bottomrule
\end{tabular}
\end{table*}

\section{Entropy Continuity for Information Parity}
\label{app:entropy-continuity}

Let $P$ and $Q$ be distributions on a $d$-point alphabet and write
$\delta=\frac12\|P-Q\|_1$.  For $0\leq\delta\leq1-1/d$, the sharp finite-
alphabet entropy-continuity inequality gives
\begin{equation}
 |H(P)-H(Q)|\leq\delta\log(d-1)+h_2(\delta)
                                                               \label{eq:entropy-cont}
\end{equation}
\cite{audenaert2007sharp}.  Marginalization cannot increase total variation,
so if two joint laws on $\cA\times\cS$ are at total variation at most
$\delta$, their $A$ and $S$ marginals are also within $\delta$.

Using $I(A;S)=H(A)+H(S)-H(A,S)$ and applying
\eqref{eq:entropy-cont} three times yields
\begin{align}
 |I_P(A;S)-I_Q(A;S)|
 \leq{}\delta\log(|\cA|-1)+\delta\log(|\cS|-1)+\delta\log(|\cA||\cS|-1)+3h_2(\delta),          \label{eq:mi-cont-detail}
\end{align}
which is~\eqref{eq:omega-def}.
In Theorem~\ref{thm:mi-parity},
let
\begin{equation}
 \widehat\delta_W:=\frac12
 \|\widehat P_{W,AS}^{(n)}-P_{W,AS}\|_1.            \label{eq:random-tv}
\end{equation}
Equation~\eqref{eq:l1-transfer}
gives
$\E\widehat\delta_W\leq\eta_n$.  Applying
\eqref{eq:mi-cont-detail} conditionally on $W$ and then Jensen's inequality
for the concave envelope of $\omega_{\cA,\cS}$ proves \eqref{eq:mi-parity-transfer}.

\section{Worked Entropy Calculations for the Discrete Models}
\label{app:examples}

This appendix verifies the source quantities used in the examples and makes
the comparison between privacy and fairness models transparent.

\subsection{Binary selective-privacy channel}

Let $X\sim\mathrm{Bernoulli}(1/2)$ and $Y=X\oplus N$ with
$N\sim\mathrm{Bernoulli}(\theta)$ independent.  Conditional on $X=x$, the
variable $Y$ equals $x$ with probability $1-\theta$ and $1-x$ with
probability $\theta$, so
\begin{equation}
 H(Y\mid X=x)=h_2(\theta),\qquad H(Y\mid X)=h_2(\theta).
                                                               \label{eq:bsc-conditional}
\end{equation}
Symmetry and the uniform input give $P_Y(0)=P_Y(1)=1/2$ and
\begin{equation}
 H(Y)=\log2,\qquad I(X;Y)=\log2-h_2(\theta).         \label{eq:bsc-information}
\end{equation}
The direct privacy threshold in~$H(Y\mid X)+\frac{\epsilon_n}{n}=H(Y)
 \quad\Longleftrightarrow\quad
 \epsilon_n=nI(X;Y)$ 
 is therefore
\begin{equation}
 \epsilon_n=n[\log2-h_2(\theta)].                   \label{eq:bsc-threshold}
\end{equation}
Below this value the residual-plus-budget branch is active; above it the
entropy cap is active. At $\theta=0$, the residual vanishes and
$I(X;Y)=\log2$.  At $\theta=1/2$, $X$ and $Y$ are independent, the residual
equals the entire task entropy, and selective privacy gives no improvement
over the entropy cap.

For the model $Y=(X,N)$ with $N\perp X$, the chain rule instead gives
\begin{align}
 H(Y\mid X)&=H(X,N\mid X)=H(N\mid X)=H(N),\nonumber\\
 H(Y)&=H(X)+H(N).                                    \label{eq:composite-task-entropy}
\end{align}
Thus the private coordinate determines only one component of the task.  The
other component is exactly the residual available for memorization under
perfect selective privacy.

\subsection{Binary fairness cascade}

In~$S\sim\mathrm{Bernoulli}(1/2),\ X=S\oplus E,
 \ T=X\oplus N,$ 
 conditionally on $X$, the task is the output of a
binary symmetric channel of crossover $\theta$, hence
$H(T\mid X)=h_2(\theta)$.  Relative to $S$, the effective error is
$E\oplus N$.  Independence gives
\begin{align}
 \Prob(E\oplus N=1)
 =q(1-\theta)+(1-q)\theta
 =q\star\theta,                                     \label{eq:binary-convolution}
\end{align}
so $H(T\mid S)=h_2(q\star\theta)$.  The sensitive input is uniform and both
binary channels are symmetric, which also makes $X$ and $T$ uniform.  Thus
\begin{align}
 H(T)&=\log2,\nonumber\\
 I(S;T)&=\log2-h_2(q\star\theta),\nonumber\\
 I(X;T)&=\log2-h_2(\theta).                          \label{eq:cascade-information}
\end{align}
The entropy cap may therefore be added explicitly to the bound in Example~\ref{ex:binary-cascade},
although it is suppressed there to emphasize the
two selective routes.

For $q,\theta\in[0,1/2]$,
\begin{equation}
 q\star\theta=\theta+q(1-2\theta)\geq\theta.         \label{eq:convolution-order}
\end{equation}
Since $h_2$ increases on $[0,1/2]$, the feature residual is no larger than the
fairness residual.  This establishes the equal-budget route comparison used
after~
$h_2(q\star\theta)-h_2(\theta)
 \leq\frac{r_n-\epsilon_n}{n}$, Section~\ref{sec:fair-route}.

\subsection{Ternary deterministic example}

For the law
\begin{equation}
 P_T=(0.6,0.25,0.15)
\end{equation}
and the map in
Example~\ref{ex:ternary-tight},
$P_S=(0.6,0.4)$. In nats,
\begin{align}
 H(T)&=-0.6\log0.6-0.25\log0.25-0.15\log0.15,
                                                               \label{eq:ternary-HT}\\
 H(S)&=-0.6\log0.6-0.4\log0.4,                     \label{eq:ternary-HS}\\
 H(T\mid S)&=H(T)-H(S),                             \label{eq:ternary-cond}
\end{align}
because $S=f(T)$.  Since $X=T$, $H(T\mid X)=0$ and $H(X\mid T)=0$.
Consequently the two direct branches are
$n[H(T)-H(S)]+\epsilon_n$ and $r_n$.  The deterministic reverse-map
conditions in Corollary~\ref{cor:fair-tight}
remove the construction penalties, so the
smaller branch is attainable in the stated regimes.

\subsection{Balanced demographic parity}

When $p_0=p_1=1/2$, each inverse-probability score
$g_s=2\ind\{S=s\}A_W(X)$ lies in $[0,2]$.  Substituting $c=2$ into
\eqref{eq:absolute-mi} gives one group-mean deviation bounded by
\begin{equation}
 2\sqrt{\frac{\kappa_n+\log2}{2n}}.
\end{equation}
Adding the two group deviations yields the factor four in $\E|\Delta_P(W)-\widehat\Delta_n(W)|
 \leq4\sqrt{\frac{\kappa_n+\log2}{2n}}.$
This calculation also shows why the gap theorem is
stated for an expected absolute error rather than merely the absolute value
of an expected signed error.

\section{Additional Rate Corollaries}
\label{app:rates}

The formulas yield several useful asymptotic summaries.  If
$\epsilon_n=O(n^a)$ with $0\leq a<1$ and $H(Y\mid X)=0$, then
\begin{equation}
 |\E[\gen]|=O\big(n^{(a-1)/2}\big).                  \label{eq:privacy-rate-a}
\end{equation}
If $H(Y\mid X)=d_n$ and $\epsilon_n=o(n)$, the selective-privacy certificate
vanishes exactly when $d_n+\epsilon_n/n\to0$ (unless a tighter compression
branch is present).  With task compression $\rho_n=O(n^b)$, $b<1$, the rate
is at most $O(n^{(b-1)/2})$ regardless of $d_n$.

For fairness, suppose
\begin{equation}
 H(T\mid S)=a_n,\quad H(T\mid X)=b_n,\quad
 \epsilon_n=O(n^c),\quad r_n=O(n^d),              \label{eq:fair-rate-setup}
\end{equation}
where $c,d<1$.  Then
\begin{equation}
 |\E[\gen]|=O\left(
 \sqrt{\min\{a_n+n^{c-1},b_n+n^{d-1}\}}
 \right).                                           \label{eq:fair-rate-summary}
\end{equation}
Thus only one conditional-entropy route must become informative.  For
fairness-metric transfer, $\kappa_n=O(n^e)$ with $e<1$ gives the parity-gap
rate $O(n^{(e-1)/2})$ when $p_{\min}$ is bounded away from zero.  The
finite-alphabet information-parity result adds the entropy-continuity modulus,
which behaves as $O(\eta_n\log(1/\eta_n))$ near zero.

For adaptive privacy with $H(Y\mid X)=0$, if the cumulative ledger satisfies
$E_{n,K}=O(n^a)$ for some $a<1$, then
\begin{equation}
 |\E[\gen]|=O\big(n^{(a-1)/2}\big).                 \label{eq:adaptive-privacy-rate}
\end{equation}
If every update spends at most $e_n$ nats, the sufficient condition is
$K_ne_n=o(n)$.  A summable schedule with
$\sum_{k\geq1}\epsilon_{n,k}=o(n)$ gives a certificate uniform over every
finite stopping time.  In adaptive fair learning, the corresponding rate is
\begin{equation}
 O\left(\sqrt{\min\left\{
 H(T\mid S)+\frac{E_{n,K}^{\rm fair}}n,
 H(T\mid X)+\frac{R_{n,K}^{\rm feat}}n,
 \frac{C_{n,K}^{\rm task}}n\right\}}\right),       \label{eq:adaptive-fair-rate}
\end{equation}
where the task branch is omitted when no direct task ledger is imposed.

\IfFileExists{IEEEtranN.bst}{\bibliographystyle{IEEEtran}}{\bibliographystyle{ieeetr}}
\bibliography{JSAIT}

\end{document}